\documentclass{SciPost}

\hypersetup{
    colorlinks,
    linkcolor={red!50!black},
    citecolor={blue!50!black},
    urlcolor={blue!80!black}
}

\usepackage[bitstream-charter]{mathdesign}
\DeclareSymbolFont{usualmathcal}{OMS}{cmsy}{m}{n}
\DeclareSymbolFontAlphabet{\mathcal}{usualmathcal}

\fancypagestyle{SPstyle}{
\fancyhf{}
\lhead{\colorbox{scipostblue}{\bf \color{white} ~SciPost Physics }}
\rhead{{\bf \color{scipostdeepblue} ~Submission }}

\fancyfoot[C]{\textbf{\thepage}}
}

\usepackage{atbegshi,picture}
\AtBeginShipoutNext{\AtBeginShipoutUpperLeft{%
  \put(\dimexpr\paperwidth-1cm\relax,-1.cm){\makebox[0pt][r]{{UWThPh 2026-10}}}%
}}

\usepackage{slashed}
\usepackage[]{cleveref}

\usepackage{IEEEtrantools}
\usepackage{braket}
\usepackage[force]{feynmp-auto}	

\definecolor{dgreen}{HTML}{008000}

\definecolor{dblue}{HTML}{0000A0}

\newcommand{\Bstrut}{\rule[-2.0ex]{0pt}{2.2ex}}

\begin{document}
\begin{fmffile}{main}
\fmfset{arrow_len}{3mm}

\pagestyle{SPstyle}

\begin{center}{\Large \textbf{\color{scipostdeepblue}{
Muon decay across scales: from LEFT to SMEFT
}}}\end{center}

\begin{center}\textbf{
Carl Jakob Moritz\textsuperscript{1} and
Tyler Corbett\textsuperscript{1$\star$}
}\end{center}

\begin{center}
{\bf 1} Universit\"at Wien
\\[\baselineskip]
$\star$ \href{mailto:corbett.t.s@gmail.com}{\small corbett.t.s@gmail.com}
\end{center}

\section*{\color{scipostdeepblue}{Abstract}}
\textbf{\boldmath{%
We revisit precision muon decay in the framework of the LEFT by considering how the LEFT perturbs the total rate as well as the standard and polarization-dependent Michel parameters. 
We derive the dependence of these observables on the dimension-five and -six LEFT operators, including lepton number violating operators, allowing for arbitrary neutrino flavor assignments. 
We consider fits of the LEFT to the muon decay data, in particular an approach to neutrino flavor general fits that does not respect the LEFT power counting and assuming new physics couples to all flavors equally and therefore allows for a consistent fit. 
We then consider how new physics imprints on the SMEFT and subsequently the LEFT in order to constrain specific scenarios of single-field extensions as well as two motivated two-field extensions. 
While we infer associated mass scales for these scenarios in the hundreds of GeV to multi-TeV range we also find that for the specific cases of lepton number violation, neutrino mass constraints from operator mixing are more restrictive than muon decay. 
}}



\noindent\rule{\textwidth}{1pt}
\tableofcontents
\noindent\rule{\textwidth}{1pt}


\section{Introduction}
\label{sec:intro}

The muon has held an important role in the development of particle physics. Following its unexpected discovery in 1936, understanding its decay properties quickly became a pathway to determining the structure of the weak interaction \cite{Steinberger:1948qba}. Michel introduced and others generalized a parameterization of the electron spectrum in muon decays \cite{Michel:1949qe,Bouchiat:1957zz,Kinoshita:1957zz,Kinoshita:1957zza}. The $\xi$ parameter, for example, is a sensitive probe of the $V-A$ form of the charged-current weak interaction. 

Since that time, there has been extensive study of the Michel parameters in the context of chiral interactions beyond the Standard Model (SM), see e.g. the muon decay review from the PDG \cite{ParticleDataGroup:2024cfk}. These studies are close precursors of modern Effective Field Theory (EFT) approaches, but do not provide the systematic power counting, operator-basis construction, renormalization-group evolution, or matching between scales supplied by the EFT approach. Indeed, many modern QFT books introduce the Fermi theory as the canonical example of an EFT using muon decay instead of  nuclear beta decay \cite{Schwartz:2014sze}.  

In the absence of signals of new physics from the LHC, EFTs have become a fundamental part of our modern approach to physics beyond the SM. They allow us to connect precision low energy physics to scales beyond the reach of modern experiments in order to better understand what new physics may exist. This is in direct analogy with the discovery of the muon and what we now understand as the connection between how it decays, through a contact four-fermion interaction, and the electroweak sector. The advantage of the EFT framework over amplitude-level parameterizations commonly employed for muon decay \cite{Fetscher:1986uj,Fetscher:2021ldh,Langacker:1988cm,TWIST:2011aa}, is that the interaction Lagrangian is organized according to a systematic expansion with a defined range of validity, while matching and renormalization-group evolution provide controlled connections between different energy scales.

In this work, we revisit muon decay in the language of the ``LEFT'' or low energy effective field theory \cite{Jenkins:2017jig} which describes physics below the electroweak scale. We connect the muon energy scale to the electroweak scale through matching\footnote{As well as running, however we find running effects to be small over these scales.}, which results in a SMEFT interpretation that connects the electroweak scale to the ultraviolet (UV) scale of some heavy new physics which hypothetically perturbs the muon decay parameters (i.e. Michel parameters). Muon decay at leading order in the SMEFT was studied in \cite{Crivellin:2021njn} for the total rate and $\eta$, while \cite{Falkowski:2015krw} added the leading SMEFT dependence for the $\beta'/A$ Michel parameter. A recent article \cite{Grunwald:2025kot} used the total decay rate to constrain part of the SMEFT parameter space in a global lepton-flavor-dependent SMEFT analysis. In contrast to the conventional treatment, we retain the full flavor structure and include lepton number violation within the LEFT. We use the conventional and electron-polarization dependent Michel parameters to constrain the parameters of the LEFT and discuss subtleties of the LEFT and SMEFT power counting where for the former lepton number violation occurs at dimension-six while for the latter at dimension-seven. While the number of free parameters affecting the Michel parameters is large for the most general LEFT approach, making assumptions about the flavor structure in the UV or choosing specific UV completions of the SM allow us to constrain many scenarios of physics beyond the SM. Notably, these constraints coming from measurements at order 100 MeV can often reach the multi-TeV range.  Tau decays have been studied in  approaches to new physics in tau decays \cite{Marquez:2022bpg}, however the corresponding Michel parameters are much less well constrained so we neglect them in our study.

Our article is organized as follows: In \Cref{sec:MichelInSM} we review muon decay in the SM and the Michel parameterization. In \Cref{sec:LEFT} we then formulate muon decay in the LEFT obtaining general expressions for each of the Michel parameters, develop our statistical treatment, and explores fits to combinations of Wilson coefficients. In \Cref{sec:weaklycoupled} we connect these low-energy results to the SMEFT and then to new physics above the electroweak scale, finding constraints on the corresponding heavy mass scales. Finally in \Cref{sec:conclusion} we summarize our results and conclude. The appendices contain extensive discussions which clarify the main text and which aid in reproducing our results.

\section{Michel Parameters}\label{sec:MichelInSM}
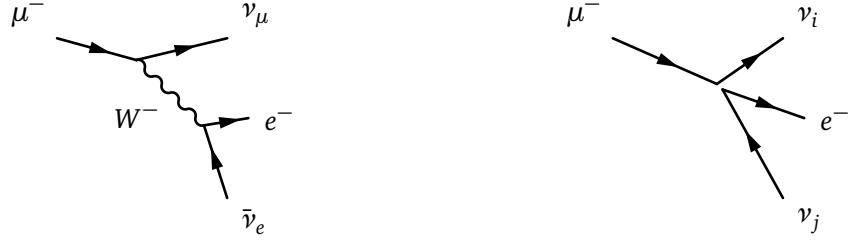
\begin{figure}
\begin{center}
\begin{tabular}{ccc}
\begin{fmfgraph*}(80,60)
	\fmfleft{l1,l2}
	\fmfright{r1,r2,r3}
	\fmf{fermion,tension=2}{l2,o1}
	\fmf{fermion}{o1,r3}
	\fmf{photon,label=$W^-$}{o1,o2}
	\fmf{fermion}{r1,o2,r2}
	\fmflabel{$\mu^-$}{l2}
	\fmflabel{$\nu_\mu$}{r3}
	\fmflabel{$e^-$}{r2}
	\fmflabel{$\bar\nu_e$}{r1}
\end{fmfgraph*}
&\qquad\qquad\qquad\qquad\qquad\qquad&
\begin{fmfgraph*}(80,60)
	\fmfleft{l1,l2}
	\fmfright{r1,r2,r3}
	\fmf{fermion,tension=2}{l2,o1}
	\fmf{phantom,tension=25}{o1,o2}
	\fmf{fermion}{o1,r3}
	\fmf{fermion}{r1,o2,r2}
	\fmflabel{$\mu^-$}{l2}
	\fmflabel{$\nu_i$}{r3}
	\fmflabel{$e^-$}{r2}
	\fmflabel{$\nu_j$}{r1}
\end{fmfgraph*}
\end{tabular}
\end{center}
\caption{Left: Muon decay in the SM proceeds via an intermediate $W$ boson. Right: In the LEFT, where the SM $W$ bosons have been integrated out of the theory, the decay proceeds via four-fermion operators. These operators can, in general, include chiral forms beyond the SM $V-A$ form. Furthermore, as the neutrinos are not detected they can also include decays to two (anti-)neutrinos of arbitrary flavor combinations.}\label{fig:Wdecay}
\end{figure}

In the Standard Model muon decay proceeds via a $W$ boson as in \Cref{fig:Wdecay}. Our momenta conventions are defined by,
\begin{equation}
\mu^-(q)\to \bar\nu_e(k_1)+\nu_\mu(k_2)+e^-(p)\, ,
\end{equation}
where $q$ denotes the muon momentum, and similarly for the other particles. As we take the neutrinos to be massless, our results are invariant under the relabelling $k_1\leftrightarrow k_2$. Later, when we consider the LEFT, lepton number violating decays into two neutrinos or antineutrinos can arise. In these cases we do not specify which $k_i$ labels which neutrino as the label is arbitrary.  More general invisible final states, including cases which can alter muon decay have also been considered in the literature \cite{Jahedi:2025hnu} including scenarios which could alter the SM Michel distribution \cite{Roig:2026wvo}.

The squared spin-summed Standard Model amplitude is given by:
\begin{align}
\left|\mathcal M\right|^2&=\frac{32}{v^4}(k_1\cdot q) (k_2\cdot p)\, ,\\
&\equiv 2^6G_{F,{\rm SM}}^2(k_1\cdot q) (k_2\cdot p)\, .
\end{align}
In the second line we have defined the Standard Model Fermi constant. Discussion of the kinematics of the decay is relegated to \Cref{sec:kinPS} to simplify our presentation here. Using these results we have for the total width in the $m_e\to 0$ limit,
\begin{equation}
\Gamma=\frac{G_{F,{\rm SM}}^2 m_\mu^5}{192\pi^3}\, .
\end{equation}
The Fermi constant is frequently taken to be an input parameter, $\hat G_F$, in SMEFT analyses. However for our analysis it provides a useful constraint, as such we choose the $\{\hat\alpha,\hat m_W,\hat m_Z\}$ input parameter scheme \cite{Crivellin:2021njn,Grunwald:2025kot}. Muon decay measurements determine $G_F$ experimentally \cite{ParticleDataGroup:2024cfk},
\begin{equation}
G_F=1.1663785(6)\cdot 10^{-5}/{\rm GeV}^2\, .\label{eq:GFexp}
\end{equation}

In the SM, we take the muon spin four vector to be $s_0=(0,P_\mu \vec n_0)$, where $P_\mu$ denotes the muon polarization, and $n_0$ is a unit vector in the spin direction. From this we obtain for the square amplitude summed over final state spins,
\begin{equation}
\left|\mathcal M\right|^2=64G_{F}^2\, p\cdot k_2\left(q\cdot k_1-m_\mu\, s_0\cdot k_1\right)\, .
\end{equation}
Next we define $W$ to be the maximum energy of the electron,
\begin{equation}
W=\frac{1}{2m_\mu}(m_\mu^2+m_e^2)\, ,
\end{equation}
and define $x=E_e/W$ which quantifies how much of the available energy is taken by the electron. The doubly differential decay width of the muon is then defined to be:
\begin{align}
\frac{d^2\Gamma}{dx\, d\cos\theta}=&\frac{m_\mu}{4\pi^3}W^4G_F^2\sqrt{x^2-x_0^2}\left[2x(1-x)+\frac{4}{9}\rho(4x^2-3x-x_0^2)+2\eta x_0(1-x)\right.\label{eq:MichelNoPe}\\
&\qquad\left.-\frac{2}{3}\xi P_\mu\cos\theta\sqrt{x^2-x_0^2}\left(\left[1-x\right]+\frac{2\delta}{3} \left[\left(4x-3\right)+\left(\sqrt{1-x_0^2}-1\right)\right]\right)\right]\, .\nonumber
\end{align}
The Standard Model values of the ``Michel parameters'' are $\rho=\delta=3/4$, $\xi=1$, and $\eta=0$. 

The Michel parameters parameterize deviations from the SM prediction within a largely model independent framework, which assumes new physics is dominated by four-fermion operators, as is the case in the LEFT at leading order. The physical interpretation of each term can be understood as follows:
\begin{itemize}
\item The term proportional to $x(1-x)$ reproduces the leading SM spectrum and reflects the underlying $V-A$ structure of the SM. Corresponds to the isotropic, or angle-independent part, of the spectrum.
\item The $\rho$ parameter isotropically modifies the overall energy dependence of the spectrum allowing for deviations from the SM. The dependence on $\rho$ is defined to vanish when integrating over the angular and energy variables.
\item The $\eta$ term isotropically controls contributions enhanced at low electron energies ($x\to0$) where electron mass effects become relevant. $\eta$ dependence remains in the total rate.
\item The $\xi$ parameter determines the extent of the parity violation in the angular distribution (anisotropic), which is maximal in the SM ($\xi=1$). The dependence on $\xi$ also drops out of the total rate when integrating over the angular and energy variables.
\item $\delta$ governs how the anisotropic contribution varies with the electron energy, $x$. It does not contribute to the total rate when integrating over the angular and energy dependence.
\end{itemize}

The experimentally extracted values are given in \Cref{tab:MichelExp}. These values are defined to be the born-level (or tree-level) parameters\cite{ParticleDataGroup:2024cfk}, i.e. the experiments have subtracted off QED radiative corrections from the data.

\begin{table}
\centering
\begin{tabular}{| r c l |  c|}
\hline
Parameter&&Measurement&Citation\\
\hline
$\rho$ &=&$0.74979\pm0.00026$&  \cite{ParticleDataGroup:2024cfk}\\
$\eta$&=&$0.057{\phantom{00}}\pm0.034$&\cite{ParticleDataGroup:2024cfk}\\
$\xi$&=&$1.0009\phantom{0}\pm0.0012$&  \cite{ParticleDataGroup:2024cfk}\\
$\delta$&=&$0.75047\pm0.00034$&  \cite{ParticleDataGroup:2024cfk}\\
$\xi'$&=&$1.00\phantom{000}\pm0.04$&\cite{ParticleDataGroup:2024cfk}\\
$\xi''$&=&$0.98\phantom{000}\pm0.04$&\cite{ParticleDataGroup:2024cfk}\\
$\eta''$&=&$0.105\phantom{00}\pm0.052$&\cite{Danneberg:2005xv}\\
$\alpha'/A$&=&$-(10\phantom{.00}\pm20)\cdot10^{-3}$&\cite{ParticleDataGroup:2024cfk}\\
$\beta'/A$&=&$\phantom{-}(\phantom{0}2\phantom{.00}\pm\phantom{0}7)\cdot10^{-3}$&\cite{ParticleDataGroup:2024cfk}\\
\hline
\end{tabular}
\caption{PDG average values for the Michel parameters. In cases where the errors in a parameter are not symmetric, we use their average. For $\eta''$ the value is taken from \cite{Danneberg:2005xv} as no value appears in the PDG. For our analyses we assume $P_\mu=1$.}\label{tab:MichelExp}
\end{table}

If the spin of the electron is also measured, additional Michel parameters can be defined: $\xi'$, $\xi''$, $\eta''$, $\alpha'/A$, and $\beta'/A$\footnote{In our construction of the Michel parameters, $A$ is not independently determined.}. We extend the parameterization of the differential decay rate as \cite{ParticleDataGroup:2024cfk,Kuno:1999jp,Marquez:2022bpg}:
\begin{equation}
\frac{d^2\Gamma}{dx\, d\cos\theta}=\frac{m_\mu }{4\pi^3} W^4G_F^2\sqrt{x^2-x_0^2}\left[F_{IS}(x)-P_\mu\cos\theta F_{AS}(x)+\hat \zeta\cdot \vec P_e(x,\theta)\right]\, .
\end{equation}
The unit vector $\hat\zeta$ defines the direction along which the electron polarization, $\vec P_e$, is analysed. From a theory perspective, we choose $\hat\zeta$ such that it projects the polarization onto a specific axis, such as longitudinal or transverse polarizations. The isotropic and anisotropic distributions correspond to those defined in \Cref{eq:MichelNoPe}\footnote{The factor of $2$ difference relative to \Cref{eq:MichelNoPe} is because we no longer sum over the electron spin states.}:
\begin{eqnarray}
F_{IS}(x)&=&x(1-x)+\frac{2}{9}\rho(4x^2-3x-x_0^2)+\eta x_0(1-x)\, ,\\
F_{AS}(x)&=&\frac{\xi}{3}\sqrt{x^2-x_0^2}\left[(1-x)+\frac{2}{3}\delta\left(\left[4x-3\right]+\left[\sqrt{1-x_0^2}-1\right]\right)\right]\, .
\end{eqnarray}
The additional electron spin dependent terms follow from expanding the electron spin dependence as $\vec P_e(x,\theta)=P_{T_1}\hat x_1+P_{T_2}\hat x_2+P_L\hat x_3$. The direction $\hat x_3$ is chosen along the electron momentum, $\hat x_1$ lies in the decay plane and is transverse to the electron momentum, and $\hat x_2$ is orthogonal to both $\hat x_1$ and $\hat x_3$. The kinematics are outlined in further detail in appendix~\ref{sec:kinPS}, our method for projecting out each Michel parameter is discussed in appendix~\ref{sec:michelbasis}. With this choice, the polarization components are given by:
\begin{IEEEeqnarray}{rl}
P_{T_1}=&\frac{P_\mu\sin\theta}{12}\left[-2\left(\xi''+12\left[\rho-\frac{3}{4}\right]\right)(1-x)x_0-3\eta\left(x^2-x_0^2\right)+\eta''\left(-3x^2+4x-x_0^2\right)\right]\, ,\nonumber\\
\\
P_{T_2}=&\frac{P_\mu\sin\theta}{3}\sqrt{x^2-x_0^2}\left[3\frac{\alpha'}{A}(1-x)+2\frac{\beta'}{A}\sqrt{1-x_0^2}\right]\, ,\\[5pt]
P_L=&-F_{IP}(x)+P_\mu\cos\theta F_{AP}(x)\, ,\\[5pt]
F_{IP}(x)=&\frac{\sqrt{x^2-x_0^2}}{54}\left[9\xi'\left(-2x+2+\sqrt{1-x_0^2}\right)+4\xi\left(\delta-\frac{3}{4}\right)\left(4x-4+\sqrt{1-x_0^2}\right)\right]\, ,\\[5pt]
F_{AP}(x)=&\frac{1}{6}\left[\xi''\left(2x^2-x-x_0^2\right)+4\left(\rho-\frac{3}{4}\right)\left(4x^2-3x-x_0^2\right)+2\eta''(1-x)x_0\right]\, .
\end{IEEEeqnarray}

\noindent These different contributions can be qualitatively understood as follows:
\begin{itemize}
\item The longitudinal polarization, $P_L$, corresponds to alignment of the electron momentum with its polarization. The corresponding contributions in addition to the usual Michel parameters are $\xi'$, $\xi''$, and $\eta''$. These terms also include contributions from $\rho$ and $\delta$ as defined above.
\item The transverse polarization $P_{T_1}$ lies in the decay plane. It receives contributions from $\xi''$ and $\eta''$ along with contributions from the previously defined $\rho$ and $\eta$. It includes angular dependence in the form of $\sin\theta$. 
\item The transverse polarization $P_{T_2}$ is perpendicular to the decay plane. It is $T$-odd and therefore sensitive to time-reversal violating effects. It also includes angular dependence of the form $\sin\theta$. The parameters $\alpha'/A$ and $\beta'/A$ are constrained by considering this kinematical region. We treat $\alpha'/A$ and $\beta'/A$ as independent variables, not ratios. Their labelling as ratios comes from a different, but equivalent, treatment of the Michel parameters.
\end{itemize}

In the SM we have for these extended Michel parameters, $\eta''=0$, $\xi'=\xi''=1$, and $\alpha'/A=\beta'/A=0$. Next we discuss the LEFT and how it modifies the Michel parameters, later in \Cref{sec:weaklycoupled} we will show how models of new physics imprint on the infrared through the SMEFT/LEFT and how the Michel parameters constrain these models.

\section{Muon decay in the LEFT}\label{sec:LEFT}

The low energy effective field theory, or LEFT, sometimes referred to as ``WET'' or weak effective theory, parameterizes physics at energy scales well below the weak scale $v\sim 246$ GeV. To properly construct the LEFT, one needs to integrate out the SM fields at the weak scale, \textit{in addition to} employing a bottom up approach to physics beyond the SM. In this manner the leading terms in the LEFT were constructed in \cite{Jenkins:2017jig}, as well as loop improved matching for the SM contribution in \cite{Dekens:2019ept,Jenkins:2017dyc}. These works also include the matching from the SMEFT on to the LEFT, which is subject to a further assumption that physics beyond the SM does not dramatically alter electroweak symmetry breaking. This has been extended to dimension seven including matching from the SMEFT in \cite{Liao:2020zyx}, and the complete dimension-eight basis which was constructed in \cite{Murphy:2020cly}.

The LEFT is expressed as:
\begin{equation}
\mathcal L=\mathcal L_{\rm renorm} + \sum_{i=1}^\infty c_i^{(4+i)}\mathcal O_i^{(4+i)}\, .
\end{equation}
Implicit in our notation is that the Wilson coefficients (or low energy constants) $c_i^{(4+i)}$ have mass dimension, $-i$, corresponding to (implicit) suppression by the heavy new physics scale $\Lambda^{-i}$. This differs from the usual treatment which takes our $c^{(4+i)}\to c^{(4+i)}/\Lambda^i$ but significantly clutters the presentation of analytic results.

The renormalizable part of the effective Lagrangian is written,
\begin{equation}
\mathcal L_{\rm renorm}=-\frac{1}{4}F_{\mu\nu}F^{\mu\nu}-\frac{1}{4}G_{\mu\nu}^AG^{A,\mu\nu}+\sum_\psi \bar\psi(i\slashed{D}-m)\psi\, ,
\end{equation}
where, for our purposes, we have dropped the $\theta$ terms and non-trivial fermion mass terms which can be found in \cite{Jenkins:2017jig}. For muon decay we will only consider operators of dimensions five and six. At dimension five we need to consider:
\begin{equation}
\mathcal L^{(5)}=c_{\nu\gamma,pr}^{(5)}\left(\nu^T_{p}C\sigma^{\mu\nu}\nu_r\right)F_{\mu\nu}+c^{(5)}_{e\gamma,pr}\left(\bar e_{L,p}\sigma^{\mu\nu}e_{R,r}\right)F_{\mu\nu}+h.c.\label{eq:L5}
\end{equation}
We drop the usual subscript ``$L$'' on neutrinos as we consider only left-handed neutrinos. The subscripts $p,r$ (and later $s,t$) are flavor indices. In order to contribute to muon decay the charged-lepton operators ($c_{e\gamma}$) necessarily correspond to flavor changing neutral currents and are therefore much better constrained by considering processes such as $\mu\to e\gamma$ \cite{MEGII:2025gzr,Calibbi:2017uvl,Pruna:2014asa,Crivellin:2014cta}. Nevertheless, we consider them in this work for completeness.

At dimension six we can break up the set of operators into lepton number preserving and violating operators. The two operator forms that preserve lepton number are hermitian\footnote{See \Cref{app:hermiticity} for a discussion of the hermiticity of these operators.} and are given by:
\begin{equation}
\mathcal L_{L{\text -}{\rm preserving}}=\left(c_{\nu e,prst}^{(6),VLL}-\frac{2}{v^2}\delta_{pt}\delta_{rs}\right) (\bar \nu_p\gamma^\mu\nu_r)(\bar e_{L,s}\gamma_\mu e_{L,t})+c_{\nu e,prst}^{(6),VLR}(\bar \nu_p\gamma^\mu\nu_r)(\bar e_{R,s}\gamma_\mu e_{R,t})\, .\label{eq:LLcons}
\end{equation}
The $2/v^2$ term is the SM contribution which we have explicitly factored out of our definition of $c_{\nu e}^{(6),VLL}$ for convenience. There are three operator forms that contribute to muon decay and violate Lepton number by two units. They are not hermitian and are given by:
\begin{align}
\mathcal L_{L{\text -}{\rm violating}}=&\, c_{\nu e,prst}^{(6),SLL}(\nu_p^T C\nu_r)(\bar e_{R,s} e_{L,t})+c_{\nu e,prst}^{(6),SLR}(\nu_p^T C\nu_r)(\bar e_{L,s}e_{R,t})\nonumber\\[5pt]
&+c_{\nu e,prst}^{(6),TLL}(\nu_p^T C\sigma^{\mu\nu}\nu_r)(\bar e_{R,s}\sigma_{\mu\nu} e_{L,t})+h.c.\label{eq:LLvio}
\end{align}

The standard approach to muon decay and the Michel parameters instead employs right handed neutrinos and the \textit{matrix element} is parameterized as follows:
\begin{equation}
\frac{4G_F}{\sqrt{2}}\sum_{\substack{\gamma=S,V,T\\ \epsilon,\eta=R,L}}g_{\epsilon\eta}^\gamma\bra{\bar e_\epsilon}\Gamma^\gamma\ket{\nu_e}\bra{\nu_\mu}\Gamma_\gamma\ket{\mu_\eta}\label{eq:MichelAmpStandard}
\end{equation}
The neutrino chirality follows from the Dirac structure ($\Gamma^\gamma$) and the chirality assigned to the charged lepton. As the neutrinos are not detected in typical muon decay experiments, the measured differential decay is insensitive to whether the underlying interaction involves right-handed neutrinos, as in \Cref{eq:MichelAmpStandard}, or charge-conjugate left-handed neutrinos as in the LEFT, \Cref{eq:L5,eq:LLcons,eq:LLvio}. Our analysis then is effectively a reinterpretation of the standard analysis in the framework of the LEFT and later SMEFT, which are absent right handed neutrinos.

Recently \cite{Breso-Pla:2025cul} used COHERENT data \cite{COHERENT:2018imc,COHERENT:2020iec,COHERENT:2021xmm} to perform a more general analysis than the typical Michel discussion mentioned in \Cref{sec:MichelInSM}. The inclusion of directly measured neutrino spectra serves to lift degeneracies in the standard Michel treatment, but does not distinguish between ``genuinely'' right-handed neutrinos, $\nu_R$ as in \Cref{eq:MichelAmpStandard}, and our LEFT treatment where we have right handed $\nu_L^C$.

\subsection{The SM-like process}\label{sec:SMlike}

We begin by deriving the Michel parameters for $\mu^-\to e^-\bar\nu_e\nu_\mu$, which corresponds to the SM process. We follow the discussion from \Cref{sec:MichelInSM} and further elaborated in appendix~\ref{sec:kinPS}. We consider strictly the SM-like process as a simple first look at our results and for the illustrative purposes of this section we will assume real Wilson coefficients. This is relaxed starting in the next section. In the next subsection we consider all four-fermion operators simultaneously which allows for different neutrino final states. Then we discuss the dipole operator case of \Cref{eq:L5}.

Starting with the contributions insensitive to the electron polarization we find:
\begin{align}
G_F^2/G_{F,{\rm SM}}^2=&1-v^2c_{\nu e,2112}^{(6),VLL}+\frac{v^4}{4}\left([c_{\nu e,2112}^{(6),VLL}]^2+[c_{\nu e,2112}^{(6),VLR}]^2\right)\nonumber\\
\rho=&\frac{3}{4}\nonumber\\
\eta=&\frac{v^2}{2}c_{\nu e,2112}^{(6),VLR}+\frac{v^4}{4}c_{\nu e,2112}^{(6),VLL}\, c_{\nu e,2112}^{(6),VLR}\label{eq:SMlikeGF}\\
\xi=&1-\frac{v^4}{2}[c_{\nu e,2112}^{(6),VLR}]^2\nonumber\\
\delta=&\frac{3}{4}\nonumber
\end{align}

As we are not yet considering the dimension-five operators of \Cref{eq:L5}, the leading contributions to these measurements go as $c^{(6)}$ i.e. $1/\Lambda^2$. Both $G_F^2$ and $\eta$ receive corrections at this order.

In \cite{Grunwald:2025kot} the authors use the $\{\hat m_W,\hat m_Z,\hat \alpha\}$ input parameter scheme. They give for the theory determination of $G_F$,
\begin{equation}
G_{F,{\rm SM}}=\frac{\pi\hat \alpha\hat m_Z^2}{\sqrt{2}\hat m_W^2(\hat m_Z^2-\hat m_W^2)}(1+\delta_R) =1.1658(10)\cdot 10^{-5}\, ,
\end{equation}
with $\delta_R=0.03686(21)$, and taking $\hat\alpha=7.2973525693(11) \cdot10^{-3}$, $\hat m_W=80.369(13)$, and $\hat m_Z=91.1880(20)$ \cite{ParticleDataGroup:2024cfk}. 
Comparing this SM prediction with the leading contribution in \Cref{eq:SMlikeGF} and the experimental value in \Cref{eq:GFexp}, we find the 95\% CL interval,
\begin{equation}
-0.004\le v^2c_{\nu e,2112}^{(6),VLL}\le 0.002\, ,
\end{equation}
taking $|c|\sim 1/\Lambda^2$ we obtain a 95\% confidence level bound on the scale of new physics:
\begin{equation}
\Lambda\ge 21v\sim {\rm\ 5.1\ TeV}
\end{equation}
Similarly if we consider the leading contribution to $\eta$ we find:
\begin{equation}
-0.019\le c_{\nu e,2112}^{(6),VLR}\,v^2\le 0.247\quad \Rightarrow\quad \Lambda\ge2v\sim 492{\rm\ GeV}\, .
\end{equation}
As this bound corresponds to the chirally ($m_e$) suppressed observable, it is not surprising the limit is much less stringent.

If we extend our analysis to allow contributions quadratic in the coefficients, and starting with the contributions insensitive to the electron polarization, we find:
\begin{align}
G_F^2/G_{F,{\rm SM}}^2=&1-v^2c_{\nu e,2112}^{(6),VLL}+\frac{v^4}{4}\left([c_{\nu e,2112}^{(6),VLL}]^2+[c_{\nu e,2112}^{(6),VLR}]^2\right)\nonumber\\
\eta=&\frac{v^2}{2}c_{\nu e,2112}^{(6),VLR}+\frac{v^4}{4}c_{\nu e,2112}^{(6),VLL}\, c_{\nu e,2112}^{(6),VLR}\label{eq:SMlikeGFquad}\\
\xi=&1-\frac{v^4}{2}[c_{\nu e,2112}^{(6),VLR}]^2\nonumber
\end{align}
$\rho$ and $\delta$ remain unchanged. Defining a simple $\chi^2$ over $G_F$, $\eta$, $\xi$, and assuming $P_\mu=1$, we find the following 95\% CL intervals for the two contributing Wilson coefficients:
\begin{equation}
\begin{array}{rcl}
-0.036\le& v^2\, c_{\nu e,2112}^{(6),VLR}&\le0.064\, ,\\
-0.0042\le& v^2\, c_{\nu e,2112}^{(6),VLL}&\le 0.0025\, .
\end{array}\label{eq:SMlikefit}
\end{equation}
These one-dimensional limits were obtained by profiling over the other parameter. In producing these limits we neglected a second minimum in the $\chi^2$ corresponding to $v^2\, c_{\nu e,2112}^{(6),VLL}\sim 4$. Some past fits in the SMEFT framework have nonetheless included such additional regions of the parameter space in fits, for example in \cite{Corbett:2015ksa,Almeida:2021asy}. For now, we argue that such a value is unphysically large. Later in our more general discussion we will incorporate the degenerate minima for completeness.  These constraints, under the assumption $|c|\sim 1/\Lambda^2$, correspond to new physics scales of $\Lambda\sim 4$ TeV and $\Lambda\sim 1$ TeV for the $LL$ and $LR$ constraints respectively. 

Extending our analysis to include electron polarization dependence we find corrections to the following additional Michel parameters:
\begin{align}
\eta''=&0-\frac{v^2}{2}c^{(6),VLR}_{\nu e,2112}-\frac{v^4}{4}c_{\nu e,2112}^{(6),VLL}c_{\nu e,2112}^{(6),VLR}=-\eta\, ,\nonumber\\
\xi'=&1-\frac{v^4}{2}[c_{\nu e,2112}^{(6),VLR}]^2=\xi\, ,\\
\end{align}
The parameters $\xi''$ and $\alpha'/A$ remain unchanged. In the LEFT we will see they only receive corrections from lepton number violating operators or dimension-five dipole operators. As the theory predictions require $\eta''=-\eta$, $\xi'=\xi$, and these two additional parameters are more poorly measured than $\eta$ and $\xi$, we neglect to extend our fit to include them here. The parameter $\beta'/A$ vanishes because of our assumption the Wilson coefficients are real, when we later relax this assumption it will receive corrections from the vector operators at linear order \cite{Falkowski:2015krw}. We will revisit these parameters in the next section where we allow for lepton flavor violation.

\subsection{Full four-fermion results}

Starting with neglecting measurements sensitive to electron spin, we find for $G_F$:
\begin{align}
G_F^2/G_{F,{\rm SM}}^2=&1-\frac{v^2}{2}\left[c_{\nu e,2112}^{(6),VLL}+c_{\nu e,1221}^{(6),VLL}\right]+\frac{v^4}{4}\sum_{a,b=1}^3\left[c_{\nu e,ab12}^{(6),VLL}c_{\nu e,ba21}^{(6),VLL}+c_{\nu e,ab12}^{(6),VLR}c_{\nu e,ba21}^{(6),VLR}\right]\nonumber\\
&+\frac{v^4}{16}\sum_{a<b}\left[|c_{\nu e,ab12}^{(6),SLL}+c_{\nu e,ba12}^{(6),SLL}|^2+|c_{\nu e,ab12}^{(6),SLR}+c_{\nu e,ba12}^{(6),SLR}|^2+(12\to 21)\right]\nonumber\\ &+2\times\frac{v^4}{16} \sum_{a=1}^{3}\left[|c_{\nu e,aa12}^{(6),SLL}|^2+|c_{\nu e,aa12}^{(6),SLR}|^2+(12\to 21)\right]\\ &+3 v^4\sum_{a<b}\left[|c_{\nu e,ab12}^{(6),TLL}-c_{\nu e,ba12}^{(6),TLL}|^2+(12\to 21)\right]\, .\nonumber \end{align}
Here we have used ``$(12\to 21)$'' to mean swapping only the last two flavor indices from ``12'' to ``21.'' For sums with $a<b$ it is understood $b$ is summed from 1 to 3. The factor of two for like neutrinos arises from the counting factor (1/2) for identical final-state particles and symmetrizing the amplitude giving $|c+c|^2$ to $4|c|^2$ for $a=b$. The factor of 3 for the tensor can be understood as coming from the six independent components of $\sigma^{\mu\nu}$ and the trace over Dirac matrices. The difference in sign structure between scalar and tensor terms is simply due to the symmetry properties of the two operators under interchange of the neutrino flavor indices.

For the other electron spin independent Michel parameters we find:
\begin{align}
\rho=&\frac{3}{4}-\frac{3v^4}{64}\sum_{a<b}\left[|c_{\nu e,ab12}^{(6),SLL}+c_{\nu e,ba12}^{(6),SLL}|^2+|c_{\nu e,ab12}^{(6),SLR}+c_{\nu e,ba12}^{(6),SLR}|^2+(12\to 21)\right]\nonumber\\
&\phantom{\frac{3}{4}}-2\times \frac{3v^4}{64}\sum_{a=1}^3\left[|c_{\nu e,aa12}^{(6),SLL}|^2+|c_{\nu e,aa12}^{(6),SLR}|^2+(12\to 21)\right]\\
&\phantom{\frac{3}{4}}+\frac{3v^4}{4}\sum_{a<b}\left[|c_{\nu e,ab12}^{(6),TLL}-c_{\nu e,ba12}^{(6),TLL}|^2+(12\to 21)\right]\, ,\nonumber\\[8pt]
\eta=&0+\frac{v^2}{4}\left[c_{\nu e,2112}^{(6),VLR}+c_{\nu e,1221}^{(6),VLR}\right]+\frac{v^4}{8}\left[c_{\nu e,1221}^{(6),VLL}c_{\nu e,2112}^{(6),VLR}+c_{\nu e,2112}^{(6),VLL}c_{\nu e,1221}^{(6),VLR}\right]\nonumber\\
&+\frac{v^4}{8}\left[c_{\nu e,2112}^{(6),VLL}c_{\nu e,2112}^{(6),VLR}+c_{\nu e,1221}^{(6),VLL}c_{\nu e,1221}^{(6),VLR}\right]-\frac{v^4}{8}\sum_{a,b=1}^3\left[c_{\nu e,ba21}^{(6),VLL}c_{\nu e,ab12}^{(6),VLR}+c_{\nu e,ab12}^{(6),VLL}c_{\nu e,ba21}^{(6),VLR}\right]\nonumber\\
&\phantom{0}+\frac{v^4}{8}\sum_{a<b}{\rm Re}\left[(c_{\nu e,ab12}^{(6),SLL}+c_{\nu e,ba12}^{(6),SLL})^*(c_{\nu e,ab12}^{(6),SLR}+c_{\nu e,ba12}^{(6),SLR})+(12\to 21)\right]\\
&\phantom{0}+\frac{v^4}{4}\sum_{a=1}^3{\rm Re}\left[(c_{\nu e,aa12}^{(6),SLL})^*(c_{\nu e,aa12}^{(6),SLR})+(12\to 21)\right]\, ,\nonumber
\end{align}

\begin{align}
\xi=&1-\frac{v^4}{2}\sum_{a,b=1}^3c_{\nu e,ab12}^{(6),VLR}c_{\nu e,ba21}^{(6),VLR}+\frac{v^4}{8}\sum_{a<b}\left[|c_{\nu e,ab12}^{(6),SLR}+c_{\nu e,ba12}^{(6),SLR}|^2-2|c_{\nu e,ab12}^{(6),SLL}+c_{\nu e,ba12}^{(6),SLL}|^2\right]\nonumber\\
&\phantom{1}+\frac{v^4}{8}\sum_{a<b}\left[-2|c_{\nu e,ab21}^{(6),SLR}+c_{\nu e,ba21}^{(6),SLR}|^2+|c_{\nu e,ab21}^{(6),SLL}+c_{\nu e,ba21}^{(6),SLL}|^2\right]\\
&\phantom{1}+\frac{v^4}{4}\sum_{a=1}^3\left[|c_{\nu e,aa21}^{(6),SLL}|^2-2|c_{\nu e,aa21}^{(6),SLR}|^2-2|c_{\nu e,aa12}^{(6),SLL}|^2+|c_{\nu e,aa12}^{(6),SLR}|^2\right]\nonumber\\
&\phantom{1}+\frac{v^4}{8}\sum_{a<b}\left[32|c_{\nu e,ab12}^{(6),TLL}-c_{\nu e,ba12}^{(6),TLL}|^2-80|c_{\nu e,ab21}^{(6),TLL}-c_{\nu e,ba21}^{(6),TLL}|^2\right]\, ,\nonumber\\
\delta=&\frac{3}{4}+\frac{9v^4}{64}\sum_{a<b}\left[|c_{\nu e,ab12}^{(6),SLL}+c_{\nu e,ba12}^{(6),SLL}|^2-|c_{\nu e,ab12}^{(6),SLR}+c_{\nu e,ba12}^{(6),SLR}|^2 -(12\to 21)\right]\nonumber\\
&\phantom{\frac{3}{4}}+\frac{9v^4}{32}\sum_{a=1}^3\left[|c_{\nu e,aa12}^{(6),SLL}|^2-|c_{\nu e,aa12}^{(6),SLR}|^2-(12\to 21)\right]\\
&\phantom{\frac{3}{4}}-\frac{9v^4}{4}\sum_{a<b}\left[|c_{\nu e,ab12}^{(6),TLL}-c_{\nu e,ba12}^{(6),TLL}|^2-(12\to 21)\right]\, .\nonumber
\end{align}
The extra quadratic term in $\eta$ that does not appear as part of a sum arises from the cross term when expanding the ratio defining eta (see \Cref{eq:solveforMichel}). It originates from the product of the linear term in the numerator ($\sim c_{\nu e,2112}^{(6),VLR}$) with the linear correction to $1/G_F^2$ ($\sim c_{\nu e,2112}^{(6),VLL}$). This contribution only exists for the SM flavor assignment. Notice that for $\xi$ there is a difference between the ``$12$'' and ``$21$'' contributions. From \Cref{eq:LLvio} we can see that exchanging these labels swaps the chirality assigned to the electron and muon. The parameter $\xi$, which corresponds to the anisotropic part of the rate, is sensitive to precisely this difference in chiral assignments. For the same reason, $\delta$ also swaps signs under the flavor exchange $(12\to 21)$.

Measurements sensitive to the electron's polarization then access the additional Michel parameters:
\begin{align}
\xi'=&1-\frac{v^4}{2}\sum_{a,b=1}^3c_{\nu e,ab12}^{(6),VLR}c_{\nu e,ba21}^{(6),VLR}-\frac{v^4}{8}\sum_{a<b}\left[|c_{\nu e,ab12}^{(6),SLL}+c_{\nu e,ba12}^{(6),SLL}|^2+|c_{\nu e,ab21}^{(6),SLR}+c_{\nu e,ba21}^{(6),SLR}|^2\right]\, ,\nonumber\\
&-\frac{v^4}{4}\sum_{a=1}^3\left[|c_{\nu e,aa12}^{(6),SLL}|^2+|c_{\nu e,aa21}^{(6),SLR}|^2\right]-6v^4\sum_{a<b}|c_{\nu e,ab12}^{(6),TLL}-c_{\nu e,ba12}^{(6),TLL}|^2\\
\xi''=&1+\frac{v^4}{8}\sum_{a<b}\left[|c_{\nu e,ab12}^{(6),SLL}+c_{\nu e,ba12}^{(6),SLL}|^2+|c_{\nu e,ab12}^{(6),SLR}+c_{\nu e,ba12}^{(6),SLR}|^2+(12\to 21)\right]\nonumber\\
&\phantom{1}+\frac{v^4}{4}\sum_{a=1}^3\left[|c_{\nu e,aa12}^{(6),SLL}|^2+|c_{\nu e,aa12}^{(6),SLR}|^2+(12\to 21)\right]\\
&\phantom{1}-10v^4\sum_{a<b}\left[|c_{\nu e,ab12}^{(6),TLL}-c_{\nu e,ba12}^{(6),TLL}|^2+(12\to 21)\right]\, ,\nonumber
\end{align}
\begin{align}
\eta''=&0-\frac{v^2}{4}\left[c_{\nu e,2112}^{(6),VLR}+c_{\nu e,1221}^{(6),VLR}\right]-\frac{v^4}{8}[c_{\nu e,1221}^{(6),VLL}c_{\nu e,2112}^{(6),VLR}+c_{\nu e,2112}^{(6),VLL}c_{\nu e,1221}^{(6),VLR}]\nonumber\\
&\phantom{0}-\frac{v^4}{8}[c_{\nu e,2112}^{(6),VLL}c_{\nu e,2112}^{(6),VLR}+c_{\nu e,1221}^{(6),VLL}c_{\nu e,1221}^{(6),VLR}]+\frac{v^4}{8}\sum_{a,b=1}^3[c_{\nu e,ba21}^{(6),VLL}c_{\nu e,ab12}^{(6),VLR}+c_{\nu e,ab12}^{(6),VLL}c_{\nu e,ba21}^{(6),VLR}]\nonumber\\
&\phantom{0}+\frac{3v^4}{8}\sum_{a<b}{\rm Re}\left[(c_{\nu e,ab12}^{(6),SLL}+c_{\nu e,ba12}^{(6),SLL})^*(c_{\nu e,ab12}^{(6),SLR}+c_{\nu e,ba12}^{(6),SLR})+(12\to 21)\right]\\
&\phantom{0}+\frac{3v^4}{4}\sum_{a=1}^3{\rm Re}\left[(c_{\nu e,aa12}^{(6),SLL})^*c_{\nu e,aa12}^{(6),SLR}+(12\to 21)\right]\, ,\nonumber\\
\frac{\alpha'}{A}=&0-\frac{v^4}{8}\sum_{a<b}{\rm Im}[(c_{\nu e,ab12}^{(6),SLL}+c_{\nu e,ba12}^{(6),SLL})^*(c_{\nu e,ab12}^{(6),SLR}+c_{\nu e,ba12}^{(6),SLR})+(12\to 21)]\nonumber\\
&\phantom{0}-\frac{v^4}{4}\sum_{a=1}^3{\rm Im}[(c_{\nu e,aa12}^{(6),SLL})^*c_{\nu e,aa12}^{(6),SLR}+(12\to 21)]\, ,\\
\frac{\beta'}{A}=&0-\frac{v^2}{4}{\rm Im}[c_{\nu e,2112}^{(6),VLR}]-\frac{v^4}{8}{\rm Im}[(c_{\nu e,2112}^{(6),VLL})^*c_{\nu e,2112}^{(6),VLR}]-\frac{v^4}{8}{\rm Im}[(c_{\nu e,2112}^{(6),VLL})c_{\nu e,2112}^{(6),VLR}]\nonumber\\
&\phantom{0}+\frac{v^4}{8}\sum_{a,b=1}^3{\rm Im}[(c_{\nu e,ab12}^{(6),VLL})^*c_{\nu e,ab12}^{(6),VLR}]\, .
\end{align}

The central values and uncertainties of the Michel parameters are given in \Cref{tab:MichelExp}. 

\subsubsection*{Dipole contributions}

The dipole contributions come from operators in Eq.~\ref{eq:L5} which introduce flavor changing neutral currents. As such we expect they are substantially better constrained by charged lepton flavor changing neutral current measurements, such as $\mu\to e\gamma$ \cite{MEGII:2025gzr,Calibbi:2017uvl,Pruna:2014asa,Crivellin:2014cta}. We include them here for completeness, however we only evaluate their contributions to $G_F$, $\rho$, $\xi$, and $\delta$, the best measured of the muon decay parameters.

We find the following results: 
\begin{align}
\frac{G_F^2}{G_{F,{\rm SM}}^2}=&1+3v^4\sum_{a<b}\left[c_{eA,12}^{(5)}(c_{eA,12}^{(5)})^*\left(c_{\nu A,ab}^{(5)}-c_{\nu A,ba}^{(5)}\right)\left(c_{\nu A,ab}^{(5)}-c_{\nu A,ba}^{(5)}\right)^*+(12\to21)\right]\nonumber\\
&\phantom{1}+3v^4\sum_{a<b}\left[c_{eA,21}^{(5)}(c_{\nu A,ab}^{(5)}-c_{\nu A,ba}^{(5)})^*(c_{\nu e,ab12}^{(6),TLL}-c_{\nu e,ba12}^{(6),TLL})+c.c.\right]\, ,\nonumber\\
\rho=&\frac{3}{4}+\frac{3v^4}{4}\sum_{a<b}\left[c_{eA,12}^{(5)}(c_{eA,12}^{(5)})^*(c_{\nu A,ab}^{(5)}-c_{\nu A,ba}^{(5)})(c_{\nu A,ab}^{(5)}-c_{\nu A,ba}^{(5)})^*+(12\to21)\right]\nonumber\\
&\phantom{{3}{4}}+\frac{3v^4}{4}\sum_{a<b}\left[c_{eA,21}^{(5)}(c_{\nu A,ab}^{(5)}-c_{\nu A,ba}^{(5)})^*(c_{\nu e,ab12}^{(6),TLL}-c_{\nu e,ba12}^{(6),TLL})+c.c.\right]\,,\nonumber
\end{align}
\begin{align}
\xi=&1-10v^4\sum_{a<b}c_{eA,12}^{(5)}(c_{eA,12}^{(5)})^*(c_{\nu A,ab}^{(5)}-c_{\nu A,ba}^{(5)})(c_{\nu A,ab}^{(5)}-c_{\nu A,ba}^{(5)})^*\\
&\phantom{1}+4v^4\sum_{a<b}c_{eA,21}^{(5)}(c_{eA,21}^{(5)})^*(c_{\nu A,ab}^{(5)}-c_{\nu A,ba}^{(5)})(c_{\nu A,ab}^{(5)}-c_{\nu A,ba}^{(5)})^*\nonumber\\
&\phantom{1}+4v^4\sum_{a<b}\left[c_{eA,21}^{(5)}(c_{\nu A,ab}^{(5)}-c_{\nu A,ba}^{(5)})^*(c_{\nu e,ab12}^{(6),TLL}-c_{\nu e,ba12}^{(6),TLL})+c.c.\right]\,,\nonumber\\
\delta=&\frac{3}{4}+\frac{9v^4}{4}\sum_{a<b}\left[c_{e A,12}^{(5)}(c_{eA,12}^{(5)})^*(c_{\nu A,ab}^{(5)}-c_{\nu A,ba}^{(5)})(c_{\nu A,ab}^{(5)}-c_{\nu A,ba}^{(5)})^*-(12\to21)\right]\nonumber\\
&\phantom{\frac{3}{4}}-\frac{9v^4}{4}\sum_{a<b}\left[c_{eA,21}^{(5)}(c_{\nu A,ab}^{(5)}-c_{\nu A,ba}^{(5)})^*(c_{\nu e,ab12}^{(6),TLL}-c_{\nu e,ba12}^{(6),TLL})+c.c.\right]\, .\nonumber
\end{align}
where ``$+c.c.$'' should be understood not to affect the flavor assignments.

\subsection{Defining our statistical approach}

In what follows, our primary approach to comparing the experimental values with our theory predictions will be a simple $\chi^2$ statistic. Here we outline the general approach which will be used throughout the remainder of our analyses. In constructing a $\chi^2$ statistic to fit the Wilson coefficients, we use the following Michel parameters,
\begin{equation}
\theta = \left( G_F^2/G_{F,{\rm SM}}^2,\,
 \rho,\,
 \delta,\,
 \xi,\,
 \eta,\,
 \eta'',\,
 \xi',\,
 \xi'',\,
 \alpha'/A,\,
 \beta'/A
 \right)\, .\label{eq:orderedmichelparams}
\end{equation}
In order to formulate a correlation matrix between these parameters:
\begin{itemize}
\item We obtain correlations among $\rho$, $\delta$, and $\xi$
from the final TWIST analysis~\cite{TWIST:2011egd}.
We have symmetrized the correlation coefficients provided by TWIST by using the arithmetic mean of these correlation coefficients, consistent with our treatment of errors in \Cref{tab:MichelExp}. 
\item The correlations between $\eta$ and $\eta''$, and separately between $\alpha'/A$ and $\beta'/A$, are taken from the general analysis of Ref.~\cite{Danneberg:2005xv}. 
\item We do not attempt to account for correlations between measurements
from different experiments, for example those arising from common
external inputs or shared systematic assumptions.
\item When no correlation information is available we set the relevant correlation to zero. This could possibly be improved through an experimental combination effort, but no such publication is available to our knowledge.
\end{itemize}
Where a PDG average is used together with a correlation coefficient
from a contributing experiment, we neglect the small modification of
the correlation induced by the additional measurements.
This approach results in the following correlation matrix,
\begin{equation}
 R =
 \begin{pmatrix}
1&0&0&0&0&0&0&0&0&0 \\
0&1&0.532&0.025&0&0&0&0&0&0\\
0&0.532&1&-0.268&0&0&0&0&0&0\\
0&0.025&-0.268&1&0&0&0&0&0&0\\
0&0&0&0&1&0.946&0&0&0&0\\
0&0&0&0&0.946&1&0&0&0&0\\
0&0&0&0&0&0&1&0&0&0\\
0&0&0&0&0&0&0&1&0&0\\
0&0&0&0&0&0&0&0&1&-0.893\\
0&0&0&0&0&0&0&0&-0.893&1
 \end{pmatrix}\, ,\label{eq:corrMatrix}
\end{equation}
where the ordering in the matrix is dictated by \Cref{eq:orderedmichelparams}.

In our subsequent analyses we form our $\chi^2$ statistic as follows. We define,
\begin{equation}
\Delta\theta_i=\left(\theta_{\rm th}(c)-\theta_{\rm exp}\right)_i\, ,
\end{equation}
where the $\theta_{\rm th}(c)_i$ represents the theory prediction of the $i^{\rm th}$ Michel parameter in \Cref{eq:orderedmichelparams} in terms of the full set of Wilson coefficients, $c$, contributing to the prediction. $(\theta_{\rm exp})_i$ is then the experimental measurement from \Cref{tab:MichelExp} as well as \Cref{eq:GFexp}. The covariance matrix is then defined as,
\begin{equation}
V_{il}=\sigma_{ij}R_{jk}\sigma_{kl}\, ,
\end{equation}
where $\sigma_{ij}$ is a diagonal matrix with elements which are the experimental error for a given experimental measurement. The $\chi^2$ is finally defined to be:
\begin{equation}
\chi^2=\Delta\theta_i (V^{-1})_{ij} \Delta\theta_j\, .\label{eq:chi2def}
\end{equation}

\subsection{Flavor general fit to combinations of Wilson coefficients}
Given the nine Michel parameters under consideration and the total rate ($G_F$) we can restrict ten independent combinations of Wilson coefficients. Unfortunately it is not possible to do so in a manner that respects the power counting. That is, the contributions linear in the Wilson coefficients cannot be differentiated from those quadratic in the coefficients without making further assumptions restricting the parameter space. The combinations which can be fit are the scalar combinations,
\begin{align}
\mathcal S_A\equiv&v^4\sum_{a<b}\left[|c_{\nu e,ab12}^{(6),SLL}+c_{\nu e,ba12}^{(6),SLL}|^2+|c_{\nu e,ab21}^{(6),SLR}+c_{\nu e,ba21}^{(6),SLR}|^2\right]+2v^4\sum_{a=1}^{3}\left[|c_{\nu e,aa12}^{(6),SLL}|^2+|c_{\nu e,aa21}^{(6),SLR}|^2\right]\,,\nonumber\\[6pt]
\mathcal S_B\equiv&v^4\sum_{a<b}\left[|c_{\nu e,ab12}^{(6),SLR}+c_{\nu e,ba12}^{(6),SLR}|^2+|c_{\nu e,ab21}^{(6),SLL}+c_{\nu e,ba21}^{(6),SLL}|^2\right]+2v^4\sum_{a=1}^{3}\left[|c_{\nu e,aa12}^{(6),SLR}|^2+|c_{\nu e,aa21}^{(6),SLL}|^2\right]\, ,\\
\mathcal Z_S\equiv&v^4\sum_{a<b}\left[(c_{\nu e,ab12}^{(6),SLL}+c_{\nu e,ba12}^{(6),SLL})^*(c_{\nu e,ab12}^{(6),SLR}+c_{\nu e,ba12}^{(6),SLR})
+(c_{\nu e,ab21}^{(6),SLL}+c_{\nu e,ba21}^{(6),SLL})^*(c_{\nu e,ab21}^{(6),SLR}+c_{\nu e,ba21}^{(6),SLR})\right]\nonumber\\
&+2v^4\sum_{a=1}^{3}\left[(c_{\nu e,aa12}^{(6),SLL})^*c_{\nu e,aa12}^{(6),SLR}+(c_{\nu e,aa21}^{(6),SLL})^*c_{\nu e,aa21}^{(6),SLR}\right]\, .\nonumber
\end{align}
The combinations $\mathcal S_A$ and $\mathcal S_B$ are sums of norms and are therefore nonnegative. $\mathcal Z_S$ is complex, and the Cauchy-Schwarz inequality relates its parts to $\mathcal S_A$ and $\mathcal S_B$ as:
\begin{equation}
{\rm Re}\left[\mathcal Z_S\right]^2+{\rm Im}\left[\mathcal Z_S\right]^2\leq\mathcal S_A\mathcal S_B\, .\label{eq:physicalConstraintZS}
\end{equation}
The tensor combinations are:
\begin{align}
\mathcal T_{12}\equiv&v^4\sum_{a<b}|
c_{\nu e,ab12}^{(6),TLL}-c_{\nu e,ba12}^{(6),TLL}|^2\, ,\\[6pt]
\mathcal T_{21}\equiv&v^4\sum_{a<b}|c_{\nu e,ab21}^{(6),TLL}-c_{\nu e,ba21}^{(6),TLL}|^2\, .\nonumber
\end{align}
Again as these are norms, we have $\mathcal T_{12}\ge0$ and $\mathcal T_{21}\ge0$. For the vector combinations we define,
\begin{align}
\mathcal V_R\equiv&v^4\sum_{a,b=1}^{3}\left|c_{\nu e,ab12}^{(6),VLR}\right|^2\, ,\nonumber\\
\mathcal G_L\equiv & -v^2{\rm Re}[c_{\nu e,2112}^{(6),VLL}]+\frac{v^4}{4}\sum_{a,b=1}^3\left|c_{\nu e,ab12}^{(6),VLL}\right|^2\, ,\nonumber\\
\mathcal E_V=&\frac{v^2}{2}{\rm Re}\left[c_{\nu e,2112}^{(6),VLR}\right]+
\frac{v^4}{4}{\rm Re}\left[(c_{\nu e,2112}^{(6),VLL})^*c_{\nu e,2112}^{(6),VLR}\right]
+\frac{v^4}{4}{\rm Re}\left[c_{\nu e,2112}^{(6),VLL}c_{\nu e,2112}^{(6),VLR}\right]
\\
&-\frac{v^4}{4}\sum_{a,b=1}^{3}{\rm Re}\left[\left(c_{\nu e,ab12}^{(6),VLL}\right)^*c_{\nu e,ab12}^{(6),VLR}\right]\, ,\nonumber\\
\mathcal B_V\equiv&-\frac{v^2}{4}{\rm Im}\left[c_{\nu e,2112}^{(6),VLR}\right]
-\frac{v^4}{8}{\rm Im}\left[(c_{\nu e,2112}^{(6),VLL})^*c_{\nu e,2112}^{(6),VLR}\right]
-\frac{v^4}{8}{\rm Im}\left[c_{\nu e,2112}^{(6),VLL}c_{\nu e,2112}^{(6),VLR}\right]
\nonumber\\
&+\frac{v^4}{8}\sum_{a,b=1}^{3}{\rm Im}\left[(c_{\nu e,ab12}^{(6),VLL})^*c_{\nu e,ab12}^{(6),VLR}\right]\, .\nonumber
\end{align}
It follows that $\mathcal V_R\ge 0$ as it is a sum of norms of Wilson coefficients. Note that $\mathcal G_L$, $\mathcal E_V$, and $\mathcal B_V$ are the terms which necessarily mix power counting in the LEFT - they depend on both terms linear and quadratic in the Wilson coefficients. 

Having defined these combinations of the parameters, the theory predictions for the Michel parameters take on a particularly simple form:
\begin{align}
G_F^2/G_{F,{\rm SM}}^2=&1+\mathcal G_L+\frac{1}{4}\mathcal V_R+\frac{1}{16}\left(\mathcal S_A+\mathcal S_B\right)+3\left(\mathcal T_{12}+\mathcal T_{21}\right)\, ,\nonumber\\[4pt]
\rho=&\frac{3}{4}-\frac{3}{64}\left(\mathcal S_A+\mathcal S_B\right)+\frac{3}{4}\left(\mathcal T_{12}+\mathcal T_{21}\right)\, ,\nonumber\\[4pt]
\delta=&\frac{3}{4}+\frac{9}{64}\left(\mathcal S_A-\mathcal S_B\right)-\frac{9}{4}\left(\mathcal T_{12}-\mathcal T_{21}\right)\, ,\nonumber\\[4pt]
\xi=&1-\frac12\mathcal V_R+\frac{1}{8}\left(\mathcal S_B-2\mathcal S_A\right)+4\mathcal T_{12}-10\mathcal T_{21}\, ,\nonumber\\[4pt]
\eta=&\mathcal E_V+\frac{1}{8}{\rm Re}[\mathcal Z_S]\, ,\\[4pt]
\eta''=&-\mathcal E_V+\frac{3}{8}{\rm Re}[\mathcal Z_S]\, ,\nonumber\\[4pt]
\xi'=&1-\frac{1}{2}\mathcal V_R-\frac{1}{8}\mathcal S_A-6\mathcal T_{12}\,,\nonumber \\[4pt]
\xi''=&1+\frac{1}{8}\left(\mathcal S_A+\mathcal S_B\right)-10\left(\mathcal T_{12}+\mathcal T_{21}\right)\, ,\nonumber\\[4pt]
\frac{\alpha'}{A}=&-\frac{1}{8}{\rm Im}[\mathcal Z_S]\, ,\nonumber\\[4pt]
\frac{\beta'}{A}=&\mathcal B_V\, .\nonumber
\end{align}

Performing a $\chi^2$ fit to the data as described above, while imposing the physical constraints such as \Cref{eq:physicalConstraintZS}, yields the following best fit values with one sigma errors:
\begin{align}
\mathcal G_L=&-0.005^{+0.007}_{-0.014}\, ,\nonumber\\
\mathcal S_A=&0.027^{+0.056}_{-0.025}\,,\nonumber\\
\mathcal E_V=&-0.006^{+0.007}_{-0.007}\,,\label{eq:boundGL}\\
{\rm Re}[\mathcal Z_S]=&0.004^{+0.011}_{-0.015}\,,\nonumber\\
{\rm Im}[\mathcal Z_S]=&0.001^{+0.012}_{-0.014}\,,\nonumber\\
\mathcal B_V=&-0.001^{+0.003}_{-0.003}\, .\nonumber
\end{align}
The following parameters hit the physical boundary for their lower bounds, with their one sigma bound imposing the upper limits:
\begin{IEEEeqnarray}{rcl}
0\le& \mathcal S_B&\le 0.004\, ,\nonumber\\
0\le& \mathcal T_{12}&\le 0.005\, ,\\
0\le& \mathcal T_{21}&\le 4\cdot10^{-5}\,,\nonumber\\
0\le& \mathcal V_R&\le 6\cdot10^{-4}\, .\nonumber
\end{IEEEeqnarray}
Their best fit points are:
\begin{align}
\left.\mathcal S_B\right|_{\rm best}=&6\cdot10^{-4}\,,\nonumber\\
\left.\mathcal T_{12}\right|_{\rm best}=&0.001\,,\\
\left.\mathcal T_{21}\right|_{\rm best}=&0\,,\nonumber\\
\left.\mathcal V_R\right|_{\rm best}=&0\, .\nonumber
\end{align}
The parameter $\mathcal T_{21}$ is better constrained than $\mathcal T_{12}$ due to a roughly degenerate direction in the space of $\mathcal S_A$, $\mathcal T_{12}$, and $\mathcal G_L$. 

A naive treatment of the parameters, $P_i$, that are quadratic in the LEFT Wilson coefficients and go as $v^4/\Lambda^4$ yields effective scales, $\Lambda_i$, scaling as,
\begin{equation}
\Lambda_i^{\rm eff}>\frac{v}{{\rm max}(|P_i^{\rm min}|,|P_i^{\rm max}|)^{1/4}}\, ,
\end{equation}
where the max/min points correspond to limits of \Cref{eq:boundGL}. Here, andfor the remainder of this article, we employ \textit{95\% CL bounds} for the scale limits, instead of the one-sigma bounds on parameters presented above.
This results in the following limits:
\begin{align}
\Lambda(\mathcal S_A)>&387\ {\rm GeV}\, ,\nonumber\\
\Lambda(\mathcal S_B)>&826\ {\rm GeV}\,,\nonumber\\
\Lambda(\mathcal T_{12})>&784\ {\rm GeV}\,,\nonumber\\
\Lambda(\mathcal T_{21})>&2,576\ {\rm GeV}\,,\\
\Lambda(\mathcal V_R)>&1304\ {\rm GeV}\,,\nonumber\\
\Lambda({\rm Re}[\mathcal Z_S])>&598\  {\rm GeV}\, ,\nonumber\\
\Lambda({\rm Im}[\mathcal Z_S])>&610\  {\rm GeV}\, .\nonumber
\end{align}
As expected for contributions largely occurring quadratically in the Wilson coefficients, we generally find weaker bounds than those found in \Cref{sec:SMlike}. 

We attempted to identify the measurements driving each bound by decomposing the change in
$\chi^2$ between the global best fit and the fully profiled
$\Delta\chi^2=1$ endpoint into contributions from the independent
experimental covariance blocks, using the individual-observable
decomposition only as a diagnostic within correlated blocks.  In particular, all of these bounds are below the 4 TeV bound found for the vector operator coupling two left-handed currents. $\mathcal T_{21}$ and $\mathcal V_R$ are both dominated (at their profiled endpoints) by the precise measurements of $\xi$ resulting in the stronger bounds. Similarly, $\mathcal S_A$ and $\mathcal T_{12}$ are predominantly constrained by $\xi'$, whose larger experimental uncertainty results in the weaker bounds. For the remaining parameters, the increase in $\chi^2$ is distributed among several observables and is affected by experimental correlations leading to a less transparent interpretation.

\subsection{Flavor democratic assumption}

If we assume that the new physics couples the same to all neutrinos, we can obtain a fit that distinguishes linear versus quadratic dependence on the Wilson coefficients. That is we make the assumption:
\begin{equation}
c_{\nu e,ab12}^{(6),X}\to c_{12}^X, \qquad c_{\nu e,ab21}^{(6),X}\to c_{21}^X\, .
\end{equation}
In this case all tensor contributions vanish exactly. This leaves us with the following set of parameters,
\begin{equation}
\mathcal G_L,\qquad \mathcal S_A,\qquad \mathcal S_B,\qquad \mathcal E_V, \qquad {\rm Re}[\mathcal Z_S],\qquad{\rm Im}[\mathcal Z_S],\qquad and\qquad \mathcal B_V\, .
\end{equation}
The only parameters which mix linear and quadratic terms are those for the vector operators. Making the following definitions,
\begin{equation}
v^2 c_{12}^{VLR}\equiv a+ib\, \qquad v^2c_{12}^{VLL}=c+id\, ,
\end{equation}
allows us to rewrite the following parameters:
\begin{align}
\mathcal V_R=&9(a^2+b^2)\ge 0\, ,\nonumber\\
\mathcal G_L =& \frac{9}{4}\left[\left(c-\frac{2}{9}\right)^2+d^2\right]-\frac{1}{9}\ge -\frac{1}{9}\,,\\
\mathcal E_V=&\frac{1}{2}a-\frac{7}{4}ac-\frac{9}{4}bd\, ,\nonumber\\
\mathcal B_V=&-\frac{1}{4}b+\frac{7}{8}bc-\frac{9}{8}ad\, .\nonumber
\end{align}
Therefore we trade the vector parameters for $a$, $b$, $c$, and $d$ which are linear order in the Wilson coefficients. The numerical factors above come from flavor counting. Rewriting our muon decay parameters in this ``democratic'' basis gives:
\begin{align}
\frac{G_F^2}{G_{F,{\rm SM}}^2}=&1-c+\frac{9}{4}(a^2+b^2+c^2+d^2)+\frac{1}{16}\left(\mathcal S_A+\mathcal S_B\right)\, ,\nonumber\\
\rho=&\frac{3}{4}-\frac{3}{64}\left(\mathcal S_A+\mathcal S_B\right)\, ,\nonumber\\
\delta=&\frac{3}{4}+\frac{9}{64}\left(\mathcal S_A-\mathcal S_B\right)\, ,\label{eq:rateuniversal}\\
\xi=&1-\frac{9}{2}(a^2+b^2)+\frac{1}{8}\left(\mathcal S_B-2\mathcal S_A\right)\, ,\nonumber\\
\eta=&\frac{1}{2}a-\frac{7}{4}ac-\frac{9}{4}bd+\frac{1}{8}{\rm Re}[\mathcal Z_S]\, ,\nonumber
\end{align}
\begin{align}
\eta''=&-\frac{1}{2}a+\frac{7}{4}ac+\frac{9}{4}bd+\frac{3}{8}{\rm Re}[\mathcal Z_S]\,,\nonumber\\
\xi'=&1-\frac{9}{2}(a^2+b^2)-\frac{1}{8}\mathcal S_A\, ,\nonumber\\
\xi''=&1+\frac{1}{8}\left(\mathcal S_A+\mathcal S_B\right)\, ,\\
\frac{\alpha'}{A}=&-\frac{1}{8}{\rm Im}[\mathcal Z_S]\,,\nonumber\\
\frac{\beta'}{A}=&-\frac{1}{4}b+\frac{7}{8}bc-\frac{9}{8}ad\, .\nonumber
\end{align}

Notice that these theory predictions result in the following relations between the theory predictions:
\begin{align}
C_1\equiv&8\left(\rho-\frac{3}{4}\right)+3(\xi''-1)=0\,,\label{eq:2parameters1}\\
C_2\equiv&8\left(\delta-\frac{3}{4}\right)+9(\xi-\xi')=0\, .\nonumber
\end{align}
These provide a simple (but non-conclusive) test of the democratic assumption. Employing the correlations and experimental values given in \Cref{eq:corrMatrix}~and~\Cref{tab:MichelExp} we can use a simple $\chi^2$ test to check the consistency,
\begin{equation}
\chi^2_{\rm identity}=\chi^2(C_1=C_2=0)-\chi^2_{\rm min}=0.27\, ,
\end{equation}
which corresponds to a $p$-value of approximately 0.88 which does not indicate tension with the democratic flavor assumption.

Performing another fit subject to our flavor assumption, we find the following one sigma bounds:
\begin{IEEEeqnarray}{rcl}
-0.011\le&a&\le0.010\, ,\nonumber\\
-0.010\le&b&\le0.011\, ,\nonumber\\
-0.002\le&c&\le 0.447\, ,\nonumber\\
-0.225\le&d&\le0.225\,,\\
0.002\le&\mathcal S_A&\le 0.006\,,\nonumber\\
0\le&\mathcal S_B&\le 0.00393\,,\nonumber\\
-0.004\le&{\rm Re}[\mathcal Z_S]&\le 0.005\,,\nonumber\\
-0.004\le&{\rm Im}[\mathcal Z_S]&\le 0.005\, ,\nonumber
\end{IEEEeqnarray}
where the lower boundary of $\mathcal S_B$ comes from the physical constraint. The best fit points, some of which are degenerate, are given by:
\begin{IEEEeqnarray}{rcl}
\left.a\right|_{\rm best}&=&(-33,-8.3)\cdot10^{-4}\,,\nonumber\\
\left.b\right|_{\rm best}&=&(-2.5,4.1)\cdot10^{-3}\,,\nonumber\\
\left.c\right|_{\rm best}&=&0.125\,,\nonumber\\
\left.d\right|_{\rm best}&=&\pm0.201\,,\\
\left.\mathcal S_A\right|_{\rm best}&=&3.9\cdot10^{-3}\,,\nonumber\\
\left.\mathcal S_B\right|_{\rm best}&=&5.0\cdot10^{-4}\,,\nonumber\\
\left.{\rm Re}[\mathcal Z_S]\right|_{\rm best}&=&1.3\cdot10^{-3}\,,\nonumber\\
\left.{\rm Im}[\mathcal Z_S]\right|_{\rm best}&=&4.3\cdot10^{-4}\, .\nonumber
\end{IEEEeqnarray}
The degeneracy is at two discrete points, the first for 
\begin{equation}
(a,b,d)=(-0.0033,-0.0025,-0.201)\,,
\end{equation}
 the second for 
 \begin{equation}
 (a,b,d)=(8.3\cdot10^{-4},0.0041,0.201)\, .
 \end{equation}

Under the democratic coupling assumption, the $\chi^2$ at the minimum rises from 6.5 for the general case to 7.4. There is no strong indication against this assumption, which is further supported by the two parameter test described below \Cref{eq:2parameters1}. It is important to stress this does not indicate, however, that the democratic assumption is favored. 

The best fit points for the scalar parameters correspond closely to those of a two measurement fit to the $\rho$ and $\delta$ parameters indicating their values are driven by these two Michel parameters. That $\mathcal S_A$ has a slightly over one-sigma discrepancy with the SM is due to the fit preferring $\rho<3/4$ and $\delta>3/4$ which moves $\mathcal S_A$ slightly away from 0. That $c$ favors exactly $1/8$ has an algebraic origin -- once $d$ is minimized according to other measurements, the total rate which is strongly constrained becomes proportional to $(c-1/8)^2$ and therefore the value $c=1/8$ is favored by the fit. The upper bound of $c$ is much higher than in \Cref{eq:SMlikefit} where we studied only the SM-like process. This is due to the degeneracy broadening the allowed range as well as summing over additional flavor contributions increasing the strength of the quadratic contribution relative to the SM-like assumption. The upper limit on $c$ corresponds closely to where the perturbative expansion has broken down. In this case $c$ is of the same order as relative factors e.g. between the linear and quadratic contribution to the total rate.

Again interpreting our bounds as 95\% CL lower limits on the mass scale of new physics gives:
\begin{eqnarray}
\Lambda(a)\ge&1.70\ {\rm TeV}\,,\nonumber\\
\Lambda(b)\ge&1.69\ {\rm TeV}\,,\nonumber\\
\Lambda(c)\ge&263\ {\rm GeV}\,,\nonumber\\
\Lambda(d)\ge&371\ {\rm GeV}\,,\\
\Lambda(\mathcal S_A)\ge&742\ {\rm GeV}\,,\nonumber\\
\Lambda(\mathcal S_B)\ge&831\ {\rm GeV}\,,\nonumber\\
\Lambda(\mathcal {\rm Re}[\mathcal Z_S])\ge&793\ {\rm GeV}\,,\nonumber\\
\Lambda(\mathcal {\rm Im}[\mathcal Z_S])\ge&800\ {\rm GeV}\,.\nonumber
\end{eqnarray}
Again, the low scale for $c$ is driven by our flavor assumptions raising the upper bound. If instead we take the lower bound $c\sim-0.002$ we obtain $\Lambda(c)\ge 5$ TeV.

\section{The SMEFT and Weakly coupled New Physics}\label{sec:weaklycoupled}
In the previous section we discussed global fits. Unfortunately, due to the number of contributing Wilson coefficients and the relatively few experimental observables, this forced us to perform a fit to aggregate variables that do not respect the power counting of the LEFT or to make broad generalizations about the flavor structure in the UV.

In this section we choose an arguably better motivated direction. First we consider strictly the directions in the LEFT space mapped to by the SMEFT matching. Then we assume a single new multiplet in the UV, integrate it out, and find the matching onto the SMEFT and subsequently onto the LEFT. From our derivation of the LEFT contributions to the muon decay parameters we can then obtain constraints on these models of new physics. Ultimately, as our observables are blind to the flavor of the neutrinos, we will still need to make flavor assumptions. Our discussion can be tailored to a specific UV model with specific flavor assumptions in a straightforward manner using the results contained in this article.

\subsection{Matching in the SMEFT}
The first complete matching of the SMEFT onto the LEFT at tree level was presented in \cite{Jenkins:2017jig}. This work dealt strictly with the dimension-five and -six SMEFT matching onto the LEFT. The authors found no matching conditions for the lepton-number violating operators in \Cref{eq:LLvio}. In \cite{Liao:2020zyx}, the authors perform the matching onto the LEFT from up to dimension-seven operators and therefore do obtain matching conditions. The following equations give the tree-level matching onto the operators relevant to our muon decay analysis:

\begin{align}
c_{\nu e,pr12}^{(6),VLL}=&-4\frac{\hat c}{\hat s}\left[c_{HWB}^{(6)}+\frac{\hat c}{4\hat s}c^{(6)}_{HD}\right]\delta^{p2}\delta^{r1}+c_{LL}^{(6),pr12}+c_{LL}^{(6),12pr}\nonumber\\
&-2\left[c_{HL}^{(6),(3)p2}\delta^{r1}+c_{HL}^{(6),(3)1r}\delta^{p2}\right]+\delta^{pr}\left[c_{HL}^{(6),(1)12}+c_{HL}^{(6),(3)12}\right]\,,\nonumber\\
c_{\nu e,pr12}^{(6),VLR}=&c_{Le}^{(6),pr12}+\delta^{pr}c_{He}^{(6),12}\,,\label{eq:matchconditionsLEFTSMEFT}\\
c_{\nu e,prst}^{(6),SLL}=&-\frac{\sqrt{2}\hat v}{8}\left(2c_{\bar eLLLH}^{(7),stpr}+c_{\bar e LLLH}^{(7),sptr}+p\leftrightarrow r\right)\,,\nonumber\\
c_{\nu e,prst}^{(6),SLR}=&-\frac{\sqrt{2}\hat v}{2}\left(c_{LeHD}^{(7),pt}\delta^{rs}+c_{LeHD}^{(7)rt}\delta^{ps}\right)\,,\nonumber\\
c_{\nu e,prst}^{(6),TLL}=&\frac{\sqrt{2}\hat v}{32}\left(c_{\bar eLLLH}^{(7),sptr}-c_{\bar eLLLH}^{(7),srtp}\right)\, .\nonumber
\end{align}
The switch from lowered flavor indices (LEFT) to raised (SMEFT) has no significance, and is just for the sake of cleaner presentation. The terms proportional to $c_{HWB}^{(6)}$ and $c_{HD}^{(6)}$ are due to our choice of the $\{\hat\alpha,\hat m_W,\hat m_Z\}$ input parameter scheme. The quantities $\hat c$, $\hat s$, and $\hat v$ are derived from the input scheme and can be found in \cite{Biekotter:2023xle}.

The SMEFT operators appearing above are given by:
\begin{align}
Q_{HWB}^{(6)}=&(H^\dagger\sigma^I H)W_{\mu\nu}^I B^{\mu\nu}\,,\nonumber\\
Q_{HD}^{(6)}=&|H^\dagger D_\mu H|^2\,,\nonumber\\
Q_{LL}^{(6),prst}=&(\bar L_p\gamma_\mu L_r)(\bar L_s\gamma^\mu L_t)\,,\nonumber\\
Q_{Le}^{(6),prst}=&(\bar L_p\gamma_\mu L_r)(\bar e_s\gamma^\mu e_t)\,,\nonumber\\
Q_{HL}^{(6),(1)pr}=&(H^\dagger i\overleftrightarrow D_\mu H)(\bar L_p\gamma^\mu L_r)\,,\\
Q_{HL}^{(6),(3)pr}=&(H^\dagger i\overleftrightarrow D^I_\mu H)(\bar L_p\sigma^I\gamma^\mu L_r)\,,\nonumber\\
Q_{He}^{(6),pr}=&(H^\dagger i\overleftrightarrow D_\mu H)(\bar e_p\gamma^\mu e_r)\,,\nonumber\\
Q_{\bar eLLLH}^{(7),prst}=&\epsilon_{ij}\epsilon_{mn}(\bar e_p L_r^i)(\overline{L_s^{C,j}}L_t^m )H^n\,,\nonumber\\
Q_{LeHD}^{(7),pr}=&\epsilon_{ij}\epsilon_{mn}(\overline{L_p^{C,i}}\gamma_\mu e_r)H^j(H^m iD^\mu H^n)\, .\nonumber
\end{align}
Here we've used $\sigma^I$ for the Pauli matrices, and the derivatives with arrows are defined as:
\begin{align}
H^\dagger i \overleftrightarrow D_\mu H=&iH^\dagger D_\mu H-i(D_\mu H)^\dagger H\,,\nonumber\\
H^\dagger i\overleftrightarrow D_\mu^I H=&iH^\dagger \sigma^I D_\mu H-i(D_\mu H)^\dagger \sigma^I H\, .
\end{align}

\Cref{tab:LEFTSMEFTscaling} shows the leading contributions to the muon decay parameters in the SMEFT. Here we have dropped all contributions except the leading SMEFT contribution. This table shows how, in the SMEFT interpretation, the LEFT power counting does not necessarily reflect the way the UV imprints on the IR for new physics that decouples well above the electroweak symmetry breaking scale. Indeed, all contributions from lepton-number violating operators' contributions to the muon decay parameters are first generated by effects of order  $\Lambda^{-6}$ in the SMEFT interpretation, which is of the same order as three insertions of dimension-six operators. 

\begin{table}
\centering
\begin{tabular}{|c |c |c |c|}
\hline
Observable&Dependence&LEFT scaling&SMEFT scaling\\
\hline
\hline
$G_F^2/G_{F,{\rm SM}}^2-1$& $c_{HWB}^{(6)},\ c_{HD}^{(6)},\ c_{LL}^{(6)},\ c_{HL}^{(6),(3)}$&$\Lambda^{-2}$&$\Lambda^{-2}$\\[5pt]
$\eta$&${\rm Re}\left[c_{Le}^{(6)}\right]$&$\Lambda^{-2}$&$\Lambda^{-2}$\\[5pt]
$\eta''$&${\rm Re}\left[c_{Le}^{(6)}\right]$&$\Lambda^{-2}$&$\Lambda^{-2}$\\[5pt]
$\beta'/A$&${\rm Im}\left[c_{Le}^{(6)}\right]$&$\Lambda^{-2}$&$\Lambda^{-2}$\\[5pt]
$\xi-1$&$\left|c_{Le}^{(6)}+c_{He}^{(6)}\right|^2$&$\Lambda^{-4}$&$\Lambda^{-4}$\\[5pt]
$\xi'-1$&$\left|c_{Le}^{(6)}+c_{He}^{(6)}\right|^2$&$\Lambda^{-4}$&$\Lambda^{-4}$\\[5pt]
$\rho-3/4$&$\left|c_{\bar eLLLH}^{(7)}\right|^2,\ \left|c_{LeHD}^{(7)}\right|^2$&$\Lambda^{-4}$&$\Lambda^{-6}$\\[5pt]
$\xi''-1$&$\left|c_{\bar eLLLH}^{(7)}\right|^2,\ \left|c_{LeHD}^{(7)}\right|^2$&$\Lambda^{-4}$&$\Lambda^{-6}$\\[5pt]
$\alpha'/A$&${\rm Im}\left[\left(c_{\bar eLLLH}^{(7)}\right)^*c_{LeHD}^{(7)}\right]$&$\Lambda^{-4}$&$\Lambda^{-6}$\\[5pt]
\hline
\end{tabular}
\caption{Muon decay observables, their leading dependence on the SMEFT, and the leading scaling according to LEFT and SMEFT power countings. The observables are ordered according to their SMEFT scaling.}\label{tab:LEFTSMEFTscaling}
\end{table}

Our treatment, in which the LEFT Lagrangian is truncated at dimension six while the resulting amplitudes are squared, does not permit a consistent fit in the SMEFT interpretation in the bottom-up approach. Dimension-six squared contributions to the matching could be incorporated, but further dimension-eight operator contributions would also be formally needed before a fit could be performed\footnote{For a partial matching to this order, see \cite{Hamoudou:2022tdn}.}. As mentioned, dimension-eight operators would also induce further angular and energy dependence not incorporated in the Michel parameters. Such a study is beyond the scope of this article, but could be interesting should future experiments have the ability to resolve higher moments in muon decay. 

In order to better frame the power of muon decay measurements in constraining new physics, the subsequent section discusses the UV matching of NP models where we can get a better handle on truncation order.

\subsection{Matching of weakly interacting NP}

Assuming some new physics occurs at a scale well-decoupled from the muon mass scale, we can integrate out the new physics and calculate its effects on the muon decay parameters. This can be achieved by integrating out a new field at some heavy mass scale, $\Lambda$, which results in matching to the SMEFT, running down to the electroweak scale, matching to the LEFT, and running down to the muon mass scale.

In the SMEFT, we checked using the \texttt{Wilson} package \cite{Aebischer:2018bkb} and \texttt{D7RGESolver} \cite{Liao:2025lxg} that the RGE effects are small. For single-operator hypotheses, running from heavy scales of order a few TeV result in corrections of order a few percent. In the LEFT, the analytic results are sufficiently simple to analyze directly, and we therefore do not make use of numerical packages. The running from the EW scale down in the LEFT results in $\mathcal O(1\%)$ effects. In the vector operator sector for the flavor assignments relevant to muon decay the one-loop, the anomalous dimensions vanish identically. The lepton number violating coefficients relevant for our discussion run multiplicatively with anomalous dimensions proportional to $\alpha_{\rm QED}$, resulting in corrections of approximately $1\text{--}3\%$. These few percent corrections are negligible compared with the assumptions we make in performing our analyses, we therefore neglect RGE effects in our analysis. Again, for a motivated UV model with specific flavor structures one could achieve better precision on the theory side (through fewer assumptions) and in this case RGE effects may become relevant. The only important exception for our analysis, is when the Weinberg operator responsible for neutrino masses is generated. We will see that neutrino mass constraints from dimension-seven SMEFT operator mixing are substantially stronger constraints than we are able to obtain from muon decay, at least for the simple models and flavor assumptions studied here.

Our matching discussion focuses on the dimension-six operator matching presented in \cite{deBlas:2017xtg} and the extended analysis up to dimension-seven from \cite{Li:2023cwy}. Our goal is to enumerate the possible single heavy multiplet extensions of the SM that generate the LEFT operators of \Cref{eq:LLcons,eq:LLvio}. We begin by identifying in \Cref{tab:UVmatching} the single heavy multiplet models that generate the SMEFT operators in \Cref{eq:matchconditionsLEFTSMEFT}.

\begin{table}
\centering
\begin{tabular}{|c| l|}
\hline
Operator & Fields\\
\hline 
\hline
$Q^{(5)}$&$S_6$, $F_1$, $F_5$\\
\hline
$Q_{HD}^{(6)}$&$S_5=(1,3)_0$, $S_6=(1,3)_1$, $V_1=(1,1)_0$, \\
& $V_2=(1,1)_1$, $V_4=(1,3)_0$\\[5pt]
$Q_{HL}^{(6),(1)}$&$F_1=(1,1)_0$, $F_2=(1,1)_{1}$, $F_5=(1,3)_0$\\
& $F_6=(1,3)_{1}$, $V_1$ \\[5pt]
$Q_{HL}^{(6),(3)}$&$F_1$, $F_2$, $F_5$, $F_6$, $V_4$ \\[5pt]
$Q_{He}^{(6)}$&$F_3=(1,2)_{1/2}$, $F_4=(1,2)_{3/2}$, $V_1$\\[5pt]
$Q_{LL}^{(6)}$ &$S_2=(1,1)_1$, $S_6$, $V_1$, $V_4$\\[5pt]
$Q_{Le}^{(6)}$&$S_4=(1,2)_{1/2}$, $V_3^\dagger=(1,2)_{-3/2}$, $V_1$\\
\hline
$Q_{\bar eLLLH}^{(7)}$&$F_1$, $F_5$\\[5pt]
$Q_{LeHD}^{(7)}$&$F_1$, $F_5$\\
\hline
\hline
\multicolumn{2}{|l|}{$S_2=\mathcal S_1$, $S_4=\varphi$, $S_5=\Xi$, $S_6=\Xi_1$,}\\
\multicolumn{2}{|l|}{$F_1=N$, $F_2=E^c$, $F_3=\Delta_1^c$, $F_4=\Delta_3^c$, $F_5=\Sigma$, $F_6=\Sigma_1^c$}\\
\multicolumn{2}{|l|}{$V_1=\mathcal B$, $V_2=\mathcal B_1$, $V_4=\mathcal W$, $V_3=\mathcal L_3^\dagger$}\\
\hline
\end{tabular}
\caption{List of SMEFT operators relevant to our muon decay analysis and which UV multiplets generate them. We also include the dimension-five Weinberg operator denoted by $Q^{(5)}$. Note that we use the notation of Li et al. \cite{Li:2023cwy} as their work more clearly indicates scalars $S$, from Fermions $F$, and vectors $V$, and includes one additional multiplet which was not relevant to the analysis in \cite{deBlas:2017xtg}. We omit the vector multiplets $\mathcal L_1$ and $\mathcal W_1$, since the derivative interactions through which they contribute are redundant and can be eliminated by field redefinitions \cite{Li:2023cwy}. At the end of the table we give the relation between the two naming schemes. The field $S_5$ contributes to $Q_{HWB}^{(6)}$ in \cite{deBlas:2017xtg}, however this is because they include dimension-five operators in their ``UV'' Lagrangians, we follow Li et al. and require the UV Lagrangian be renormalizable.}\label{tab:UVmatching}
\end{table}

From \Cref{tab:UVmatching}, we immediately see that neutrino masses are generated at tree level by the $S_6$, $F_1$, and $F_5$ models. This conclusion can be evaded in extended, symmetry-protected seesaw constructions such as the inverse seesaw, where an approximate lepton-number symmetry suppresses the light-neutrino masses while allowing sizable active--heavy mixing at comparatively low scales \cite{Forero:2011pc}. The corresponding dimension-five coefficient can therefore be suppressed while lepton-number-conserving dimension-six effects remain appreciable\cite{Broncano:2002rw}. We begin by assuming that, for these models, such a suppression exists. As the contributions to the dimension-seven operators are proportional to those of the Weinberg operator, this assumption also removes their contributions.

Under the assumptions outlined above, we can largely immediately constrain the theories directly from muon decay. In some cases we have two different parameters contributing to the same SMEFT/LEFT coefficients. For example in the $S_6$ model, we have contributions to $c_{\nu e,pr12}^{(6),VLL}$ of \Cref{eq:matchconditionsLEFTSMEFT} coming from two different SMEFT operators with different coupling dependence. In this model, we have,
\begin{align}
c_{HD}^{(6)}=& \frac{2|\mu|^2}{M^4}\,,\\
c_{LL}^{(6),2112}+c_{LL}^{(6),1221}\sim &\frac{|Y_{12}+Y_{21}|^2}{4M^2}\, ,\nonumber
\end{align}
where $\mu$ is a dimension-one trilinear coupling,  $Y$ is a new Yukawa coupling, and $M$ is the heavy mass of the $S_6$. The LEFT coefficient then becomes:
\begin{align}
c_{\nu e,2112}^{(6),VLL}=-\frac{2\hat c^2}{\hat s^2}\frac{|\mu|^2}{M^4}+\frac{|Y_{12}+Y_{21}|^2}{M^2}\, .
\end{align}
In our analyses we will make the assumption that the terms going as $1/M^2$ dominate which corresponds to a further assumption about the relationship between the new dimensionful coupling and the new Yukawa coupling: $\mu^2/M^2\ll Y^2$. These details can be ascertained from \Cref{tab:UVmatching_scaling} which contains the information about the matching subject to the assumptions about structures discussed above.

\begin{table}
\centering
\renewcommand{\arraystretch}{1.7}
\begin{tabular}{|l|l|c|}
\hline
UV model & Muon-decay-relevant dimension-six SMEFT matching&$\mathbb{R}$ or $\mathbb C^*$\\
\hline
\hline
$S_2=(1,1)_1$\Bstrut&$\displaystyle c_{LL}^{(6)} \sim -\frac{Y Y^\ast}{M^2}$&$\mathbb R$\\
\hline
$S_4=(1,2)_{1/2}$\Bstrut&$\displaystyle c_{Le}^{(6)} \sim -\frac{Y Y^\ast}{M^2}$&$\mathbb C$\\
\hline
$S_5=(1,3)_0$\Bstrut&$\displaystyle c_{HD}^{(6)} \sim -\frac{|\mu|^2}{M^4}$&$\mathbb R$\\
\hline
$S_6=(1,3)_1$\Bstrut&$\displaystyle c_{HD}^{(6)} \sim \frac{|\mu|^2}{M^4},\quad c_{LL}^{(6)} \sim \frac{Y Y^\ast}{M^2}$&$\mathbb R$\\
\hline
$F_1=(1,1)_0$\Bstrut&$\displaystyle c_{HL}^{(6),(1)} \sim \frac{Y Y^\ast}{4M^2},\quad c_{HL}^{(6),(3)} \sim -\frac{Y Y^\ast}{4M^2}$&$\mathbb R$
\\
\hline
$F_2=(1,1)_1$\Bstrut&$\displaystyle c_{HL}^{(6),(1)} \sim -\frac{Y Y^\ast}{4M^2}, \quad c_{HL}^{(6),(3)} \sim -\frac{Y Y^\ast}{4M^2}$&$\mathbb R$\\
\hline
$F_3=(1,2)_{1/2}$\Bstrut&$\displaystyle c_{He}^{(6)} \sim \frac{Y Y^\ast}{M^2}$&--\\
\hline
$F_4=(1,2)_{3/2}$\Bstrut&$c_{He}^{(6)} \sim -\dfrac{Y Y^\ast}{M^2}$&--\\
\hline
$F_5=(1,3)_0$\Bstrut&$\displaystyle c_{HL}^{(6),(1)} \sim \frac{3Y Y^\ast}{8M^2},\quad c_{HL}^{(6),(3)} \sim \frac{Y Y^\ast}{8M^2}$&$\mathbb R$\\
\hline
$F_6=(1,3)_1$\Bstrut&$\displaystyle c_{HL}^{(6),(1)} \sim -\frac{3Y Y^\ast}{8M^2},\quad c_{HL}^{(6),(3)} \sim +\frac{Y Y^\ast}{8M^2}$&$\mathbb R$\\
\hline
$V_1=(1,1)_0$\Bstrut&$\displaystyle c_{HD}^{(6)} \sim -2\frac{g_H^2}{M^2},\quad c_{HL}^{(6),(1)} \sim -\frac{g_H g_L}{M^2}, \quad c_{He}^{(6)} \sim -\frac{g_H g_e}{M^2},$&mixed\\
&$ \quad c_{LL}^{(6)} \sim -\frac{g_L g_L}{2M^2}, \quad c_{Le}^{(6)} \sim -\frac{g_L g_e}{M^2} $& \\
\hline
$V_2=(1,1)_1$\Bstrut & $c_{HD}^{(6)} \sim 4\dfrac{|g|^2}{M^2}$&$\mathbb R$ \\
\hline
$V_3=(1,2)_{3/2}$\Bstrut & $c_{Le}^{(6)} \sim \dfrac{g g^\ast}{M^2}$&$\mathbb C$ \\
\hline
$V_4=(1,3)_0$\Bstrut & $\displaystyle c_{HD}^{(6)} \sim \frac{g_H^2}{M^2}, \quad c_{HL}^{(6),(3)} \sim \frac{g_H g_L}{M^2}, \quad c_{LL}^{(6)} \sim \frac{g_L g_L}{M^2} $&$\mathbb R$
\\
\hline
\end{tabular}
\caption{Parametric tree-level matching of the UV multiplets onto the
dimension-six SMEFT operators relevant to muon decay. Flavor indices and
numerical matching coefficients are suppressed. Where multiple operators are generated and they have the same coupling structures, the correlations between the operators presented in the table are exact. E.g. for $F_5$ the ratio between the two generated operators is 3, even when our assumptions simplifying flavor structures are not made. This does not hold for $V_4$ where the coupling structures differ as the $V_4$ couples to different multiplets with different strengths ($g_i$). Gauge-like couplings are denoted $g$ or $g_i$ when more than one coupling is present, Yukawa couplings as $Y$, and trilinear scalar couplings as $\mu$. $^*$The $\mathbb{R}$ or $\mathbb C$ column indicates if the coefficients are real or complex \textit{for neutrinos with the SM-flavor assignment.}}
\label{tab:UVmatching_scaling}
\end{table}

\subsubsection*{SM-like neutrino assumption}
We begin by assuming that the NP only couples to SM-like neutrino assignments. This immediately removes the dependence on $c_{HL}^{(6),(1)}$ and $c_{He}^{(6)}$ as they require like final-state neutrinos. As shown in the last column of \Cref{tab:UVmatching_scaling} this necessarily makes the coupling combinations appearing in the Wilson coefficients real. We fit each model separately and find bounds on the coupling-to-mass ratios. 

We start with the models which only shift the vector left-handed LEFT operator couplings, i.e. those that generate $c_{LL}^{(6)}$, $c_{HD}^{(6)}$, or $c_{HL}^{(6),(3)}$. First considering $S_2$ we find for a linear dimension-six fit,
\begin{equation}
v^2c_{LL}^{(6)}=(-0.1^{+0.1}_{-1.7})\cdot 10^{-3}\, .
\end{equation}
As $c_{LL}^{(6)}$ is constrained to be strictly negative for this model, under our flavor assumptions, the upper bound is identically zero. In terms of 95\% confidence level bounds on the new physics we have:
\begin{align}
\frac{M}{|Y|}\gtrsim 3.7{\rm\ TeV}\, .
\end{align}
The difference between this bound and the one obtained in \Cref{sec:SMlike} is due to the requirement the effective coupling be negative. This model, matching, and effect on muon decay was studied in detail in \cite{Crivellin:2020klg}.

Rescaling this result by $4\hat c^2/\hat s^2$ gives the bound for the $V_2$ model:
\begin{align}
\frac{M}{|g|}\gtrsim 12{\rm\ TeV}\, .
\end{align}
Similarly we find for the mass bounds for both $F_5$ and $F_6$:
\begin{align}
\frac{M}{|Y|}\gtrsim 1.3{\rm\ TeV}\, .
\end{align}
For $S_5$ the sign of $c_{HD}^{(6)}$ is results in a best fit point with one-sigma errors:
\begin{equation}
v^2 c_{HD}^{(6)}=(0^{+0}_{-2.8})\cdot 10^{-4}\, ,
\end{equation}
and the resulting 95\% CL limit on the mass-coupling ratio is:
\begin{equation}
\frac{M^2}{|\mu|}\gtrsim 9.2{\rm\ TeV}\, .
\end{equation}
Rescaling this result also gives us the limit on the $S_6$ model (subject to the assumption\\
 $|\mu|^2/M^2\ll |Y|^2$):
\begin{equation}
\frac{M}{|Y|}\gtrsim 2.5{\rm\ TeV}\, .
\end{equation}
And similarly for the $F_1$ and $F_2$ models we find:
\begin{equation}
\frac{M}{|Y|}\gtrsim 3.5{\rm\ TeV}
\end{equation}
For $V_4$ the sign of the Wilson coefficient is not fixed, and we instead find for our interval:
\begin{equation}
v^2c_{LL}^{(6)}=(-0.1\pm 1.7)\cdot 10^{-3}
\end{equation}
From this we instead constrain a combination of gauge-like couplings. Defining,
\begin{equation}
K_{V_4}=g_L^2-2g_Hg_L-\frac{\hat c^2}{\hat s^2}g_H^2\,,
\end{equation}
we obtain the limit:
\begin{equation}
\frac{M}{\sqrt{|K_{V_4}|}}\gtrsim 3.7 {\rm\ TeV}\, .
\end{equation}

In all of the above cases, to the quoted precision, the dimension-six and dimension-six-squared fits are indistinguishable. Our bounds above arise entirely from $G_F$, or equivalently the total muon decay rate. In Ref.~\cite{Ellis:2020unq}, the authors instead employ the $\{\hat\alpha,\hat G_F,\hat m_Z\}$ input scheme where this sensitivity is redistributed among the remaining electroweak observables. Their fit is to $Z$-pole, electroweak, and other LHC observables. A coefficient-by-coefficient comparison is therefore not strictly like-for-like, but it nevertheless provides a useful indication of the relative sensitivity of muon decay as compared with the broader electroweak fit. For $c_{HWB}^{(6)}$, $c_{HD}^{(6)}$, and $c_{HL}^{(3)}$, our single-operator limits are generally of the same order as those of \cite{Ellis:2020unq}, with individual endpoints typically weaker by factors of 2-8. The main exception being $c_{LL}^{(6)}$ where our positive and negative limits are weaker by factors of approximately $2$ and 14, respectively. Conversely, our negative bound of $c_{HD}^{(6)}$ is approximately a factor of two stronger.

Next we consider theories that strictly generate the right-handed coupling to charged leptons. As such, our fits incorporate many muon decay parameters: $G_F$, $\eta^{(\prime\prime)}$, $\xi^{(\prime)}$, and $\beta'/A$. 
Fitting the single complex parameter $c_{Le}^{(6)}$ we obtain for the linear fit:
\begin{align}
v^2{\rm Re}[c_{Le}^{(6)}]=&(-1.3\pm 1.4)\cdot10^{-2}\, ,\nonumber\\
v^2{\rm Im}[c_{Le}^{(6)}]=&(0.5\pm 1.3)\cdot10^{-2}\, ,
\end{align}
Including dimension-six squared contributions gives slightly more restrictive bounds:
\begin{align}
v^2{\rm Re}[c_{Le}^{(6)}]=&(-1.1\pm 1.2)\cdot10^{-2}\, ,\nonumber\\
v^2{\rm Im}[c_{Le}^{(6)}]=&(0.4\pm 1.1)\cdot10^{-2}\, .
\end{align}
We identify the largest value of the complex coefficient allowed by the fit in order to obtain limits on the mass-to-coupling ratios for the $S_4$ and $V_3$ models finding, for the dimension-six squared fit:
\begin{align}
\frac{\Lambda(S_4)}{\sqrt{|Y_{22}Y_{11}|}}\gtrsim& 870{\rm\ GeV}\,,\nonumber\\
\frac{\Lambda(V_3)}{\sqrt{|g_{12}g_{21}|}}\gtrsim &1.2{\rm\ TeV}\,.
\end{align}

The case of $V_1$ is the most complicated as it contributes to the $LL$ and $LR$ currents. In this case we can bound three independent combinations of UV parameters. We find,
\begin{align}
v^2K_{V_1}/M^2\equiv v^2\left(2\frac{\hat c^2}{\hat s^2}g_H^2-g_L^2\right)/M^2=&(-0.1\pm1.7)\cdot 10^{-3}\nonumber\,,\\
v^2{\rm Re}[g_L^* g_e]/M^2=&(1.0\pm 1.2)\cdot10^{-2}\, ,\\
v^2{\rm Im}[g_L^* g_e]/M^2=&(0.3\pm1.1)\cdot10^{-2}\nonumber\, .
\end{align}
These correspond to 95\% CL bounds on the mass-coupling ratios:
\begin{align}
\frac{M}{\sqrt{|K_{V_1}|}}\gtrsim 3.7{\rm\ TeV}\,,\nonumber\\
\frac{M}{\sqrt{|g_L^* g_e|}}\gtrsim 1.2{\rm\ TeV}\,.
\end{align}

\subsubsection*{Relaxing the flavor assumption}

We can, through some further assumptions, relax the SM-like flavor assumption. This only works for the case of $S_4$ and $V_3$ as they generate $c_{Le}^{(6)}$ which is constrained by multiple muon decay parameters. For the models that generate $LL$ currents, only the total rate constrains the parameters of the model and we cannot learn any additional useful information. In the case of $V_1$ we do not have enough constraints to perform a similar analysis without further assumptions.

Taking $S_4$ as a working example, we will assume that the UV model couples to a subset of neutrinos with the same strength as to the SM-like flavor assignment. The matching conditions for the flavor general case are:
\begin{equation}
c_{Le}^{(6),prst}=-\frac{Y_{tp}Y_{sr}^*}{M^2}\, .
\end{equation}
We will assume that the NP couples to $N_{\rm ex}$ flavor combinations of neutrinos with the same strength as to the SM assignment. This gives for the sum over flavors,
\begin{equation}
\sum_{p,r}\left|c_{Le}^{(6),pr12}\right|^2=(1+N_{\rm ex})\left|c_{Le}^{(6),2112}\right|^2\, .
\end{equation}
As we have linear and quadratic contributions we can then distinguish $N_{\rm ex}$ from $c_{Le}^{(6),2112}$. We limit our consideration to up to $8$ additional combinations of final state flavors of neutrinos, the maximum number allowed by the SMEFT assumption\footnote{The SMEFT can be adapted to addition neutrinos, the most common case being light right-handed sterile Dirac neutrinos as in the $\nu$SMEFT\cite{Liao:2016qyd,Li:2021tsq}.}. We treat $N_{\rm ex}$ as a continuous variable, as such it is an effective number of additional final state combinations of neutrinos. We then perform a two-dimensional fit in the parameters $N_{\rm ex}$ and 
\begin{equation}
\Lambda_{S_4}=\frac{M}{\sqrt{|Y_{22}Y_{11}|}}\, .
\end{equation}
The best fit point is located at $N_{\rm ex}=0$ with a value $\Lambda_{S_4}\sim1.6$ TeV. 

\begin{figure}
\includegraphics{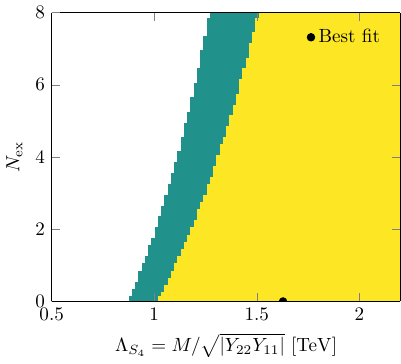}\includegraphics{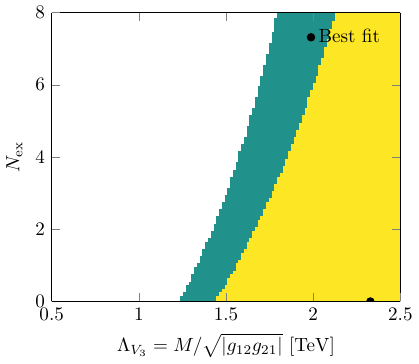}
\caption{2D constraints on $\Lambda$ vs. $N_{\rm ex}$. The dark green (left-most) region corresponds to 95\% CL, while the yellow (right-most) to 68\% CL regions. The best fit points prefer $N_{\rm ex}=0$. For lower $\Lambda$ the fit is able to restrict the number of contributing flavors, subject to our assumptions as discussed in the text. As $\Lambda$ is raised the allowed $N_{\rm ex}$ quickly exceeds the number of flavor combinations allowed under the SMEFT assumption.}\label{fig:NexLambdaplots}
\end{figure}

\Cref{fig:NexLambdaplots} (left) shows the 68\% and 95\% confidence regions.  While the number of additional neutrino combinations contributing is not bounded for sufficiently large $\Lambda$, smaller values do require a more limited number of participating neutrinos. We reiterate that this is subject to our assumption that all neutrinos couple with the same strength and is therefore an \textit{effective} number of neutrino combinations. We repeat the same exercise for $V_3$, the matching in this case simply involves an additional factor of 2, which results in bounds shifted by a factor of $\sqrt{2}$. In this case, the best fit point is again at $N_{\rm ex}=0$ with a value $\Lambda_{V_3}\sim 2.3$ TeV. \Cref{fig:NexLambdaplots} (right) shows the corresponding two-dimensional plot for this model.

\subsubsection*{Two-field extensions}
If we relax the single field assumption we can avoid generating the Weinberg operator at tree level. We consider two such models from \cite{Li:2023cwy}, they are chosen as their parameter dependence in the IR is particularly simple allowing for a straightforward fit. Again we will make the assumption of strictly the SM neutrino assignment in the final state, otherwise the fit does not close.

The first model we consider includes the $S_2$ and $F_4$. The part of the UV Lagrangian relevant to muon decay can be written as,
\begin{equation}
\Delta\mathcal L = -Y_{LL}^{rs}\epsilon^{ij}(L_{ir} C L_{js})S_2+Y_{eF}^r(\bar e_r C \bar F_4^i) H_i-Y_{FL}^s\epsilon^{ij}(F_{4i} C L_{js})S_2^\dagger + h.c.
\end{equation}
This results in the matching onto the IR,
\begin{align}
c_{LL}^{(6),prst}=&-\frac{Y_{LL}^{pr}(Y_{LL}^*)^{ts}}{M_{S_2}^2}\nonumber\,,\\
c_{He}^{(6),pr}=&-\frac{Y_{eF}^{p}(Y_{eF}^*)^r}{2M_{F_4}^2}\,,\\
c_{\bar eLLLH}^{(7),prst}=&\frac{2Y_{eF}^p Y_{FL}^t Y_{LL}^{rs}}{M_{F_4}M_{S_2}^2}\nonumber\, .
\end{align}
Imposing that the final state neutrinos have the SM assignments then restricts us to a subspace where $c_{He}^{(6)}$ does not contribute to the muon decay process and we find the following coefficients which do contribute:
\begin{align}
c_{LL}^{(6),2112}=c_{LL}^{(6),1221}=&-\frac{|Y_{LL}^{12}|^2}{M_{S_2}^2}\nonumber\,,\\
c_{\bar eLLLH}^{(7),1122}=&\frac{2Y_{eF}^1Y_{FL}^2Y_{LL}^{12}}{M_{F_4}M_{S_2}^2}\label{eq:matchingS2F4}\,,\\
c_{\bar eLLLH}^{(7),1212}=&-\frac{2Y_{eF}^1Y_{FL}^2Y_{LL}^{12}}{M_{F_4}M_{S_2}^2}\nonumber\,.
\end{align}
In order to establish a meaningful power counting, we truncate our calculations of the amplitudes at dimension-seven and square. While formally inconsistent -- we are including incomplete contributions to observables up to order $\Lambda^{-6}$ while neglecting amplitude level contributions of lower order -- our approach is consistent with one of the guidelines outlined by the LHC EFT WG \cite{Brivio:2022pyi}. We only find contributions to $G_F$, $\xi'$, and $\xi''$. Contributions from $c_{\bar eLLLH}^{(7)}$ to $\rho$, $\xi$, and $\delta$ vanish identically due to our flavor assumptions (see \Cref{eq:matchconditionsLEFTSMEFT}).

Fitting this model to the muon decay measurements results in the following constraints,
 \begin{align}
 v^2c_{LL}^{(6),2112}=&(0^{+0}_{-1.3})\cdot10^{-4}\,,\\
 v^3\left|c_{\bar eLLLH}^{(7),1122}\right|=&(1.8^{+1.1}_{-1.8})\cdot10^{-1}\, .
 \end{align}
$c_{LL}^{(6)}$ is physically restricted to be negative by the matching condition in \Cref{eq:matchingS2F4}, while only the magnitude of the dimension-seven Wilson coefficient contributes to muon decay. The best fit mildly favors a nonzero $c_{\bar eLLLH}^{(7)}$ as its quadratic contribution both increases the total rate and lowers $\xi'$ and $\xi''$. Lowering $\xi''$ improves agreement with experiment while lowering $\xi'$ worsens agreement. The overall significance appears overstated by our way of expressing the lower error (which is the boundary), and corresponds only to $\Delta\chi^2\sim0.05$.

\begin{figure}
\centering
\includegraphics{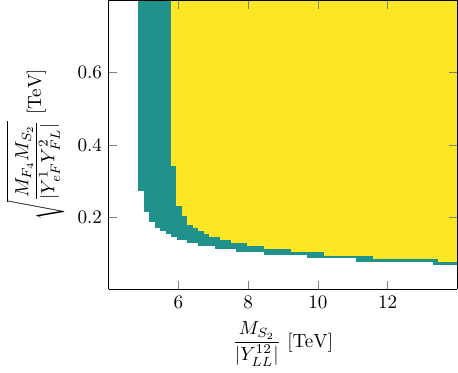}\includegraphics{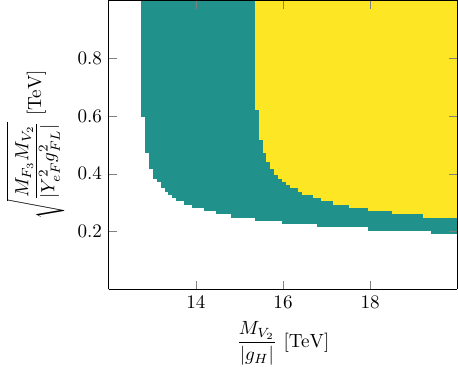}
\caption{68\% (yellow) and 95\% (dark green) confidence level regions for the two mass-to-parameter ratios shown. The left plot is for the $S_2$--$F_4$ model, while the right is for the $F_3$--$V_2$ model.}\label{fig:2DMasses}
\end{figure}

Converting to the UV parameters we obtain 95\% CL bounds on the ratio of mass to the parameters,
\begin{align}
\frac{M_{S_2}}{|Y_{LL}^{12}|}\gtrsim& 5.43{\rm\ TeV}\,,\label{eq:masslimits2field1}\\
\sqrt[3]{\frac{M_{F_4}M_{S_2}^2}{|Y_{eF}^1Y_{FL}^2 Y_{LL}^{12}|}}\gtrsim& 430{\rm\ GeV}\, .\nonumber
\end{align}
\Cref{fig:2DMasses} shows the 68\% and 95\% confidence level regions for two mass ratios. A mild increase in $M_{S_2}/|Y_{LL}^{12}|$ to around 7 TeV leaves the other ratio effectively unbounded from above. This is, naturally, expected as this decoupling behavior is built into the SMEFT matching assumption.

As another example, we consider a model with $F_3$ and $V_2$ fields. In this case the relevant part of the UV Lagrangian is given by,
\begin{equation}
\Delta\mathcal L=-Y_{eF}^r \epsilon_{ij}(\bar e_r F_3^i)H^{\dagger j}-2g_H\epsilon_{ij}(D_\mu H)^i H^j V_2^{\dagger\mu}+g_{FL}^s(\bar L_s\gamma_\mu F_{3i})V_2^{\dagger\mu}+h.c.
\end{equation}
The resulting matching onto the relevant parts of the SMEFT is given by,
\begin{align}
c_{He}^{(6),pr}=&\frac{Y_{eF}^p(Y_{eF}^r)^*}{2M_{F_3}^2}\,,\nonumber\\
c_{HD}^{(6)}=&\frac{4|g_H|^2}{M_{V_2}^2}\,,\\
c_{LeHD}^{(7),pr}=&\frac{2i(Y_{eF}^r)^*(g_{FL}^p)^*g_H}{M_{F_3}M_{V_2}^2}\, .\nonumber
\end{align}
We note that $c_{HD}$ is positive definite. Imposing that the new physics only couples to the SM-like neutrino assignment again causes $c_{He}^{(6)}$ not to contribute, $c_{HD}^{(6)}$ is flavor independent and so does not change, and the dimension-seven operator coefficient becomes,
\begin{equation}
c_{LeHD}^{(7),22}=\frac{2i(Y_{eF}^2)^*(g_{FL}^2)^* g_H}{M_{F_3}M_{V_2}^2}\, ,
\end{equation}
where the superscripts ``2'' in the numerator should be understood to be flavor labels, not squares. In this case the model only contributes to the LEFT coefficient $c_{\nu e}^{(6),SLR}$. Under our flavor assumptions, this model only shifts $G_F$, $\rho$, $\xi$, $\delta$, and $\xi''$.

Fitting the two relevant SMEFT coefficients gives,
\begin{align}
v^2c^{(6)}_{HD}=&(2.9^{+4.9}_{-2.9})\cdot 10^{-4}\,,\nonumber\\
v^3|c_{LeHD}^{(7),22}|=&(0^{+1.8}_{-0})\cdot10^{-2}\, .
\end{align}
Converting these to bounds on the relevant masses and parameters, we obtain the 95\% CL regions:
\begin{align}
\frac{M_{V_2}}{|g_H|}\gtrsim& 14{\rm\ TeV}\, ,\label{eq:masslimits2field2}\\
\sqrt[3]{\frac{M_{F_3}M_{V_2}^2}{|Y_{eF}^2 g_{FL}^2 g_H|}}\gtrsim& 966{\rm\ GeV}\nonumber
\end{align}
As with the last model, two-dimensional contours for the lower bound on the mass-parameter ratios can be found in \Cref{fig:2DMasses}. Again, we find that minor adjustment to the UV parameters allows for the UV scale corresponding to $M_{F_3}$ to be taken arbitrarily high.

However, if we take into account that these lepton-number violating operators generate the Weinberg operator through running we find that neutrino mass constraints are substantially stronger constraints on these models. For the mass-coupling limits we found in \Cref{eq:masslimits2field1,eq:masslimits2field2} we find neutrino masses from 100 to 10,000 eV. At least for our simple models, subject to our flavor assumptions, it appears that muon decay is not provide competitive sensitivity to the lepton-number violating operators.

\section{Conclusions}\label{sec:conclusion}
In this work we revisited muon decay and reinterpreted the muon decay measurements in the language of EFTs. We began by reviewing the standard muon decay parameterization which included the total rate, the standard ``Michel'' parameters $\rho$, $\xi$, $\delta$, and $\eta$, as well as the extended set depending on electron polarization dependent measurements. 

Next, we briefly discussed the formulation of the LEFT which is relevant at the scale of the muon, elaborated the operators which could contribute to muon decay, and calculated each of the muon decay parameters. By adopting the LEFT interpretation our final state neutrinos were fixed to be left-handed, contrary to e.g. the PDG treatment \cite{ParticleDataGroup:2024cfk} where their chirality is not fixed, and we summed all possible flavor combinations which could contribute including lepton-number violating final states. In this discussion we included dipole operators for completeness, but limited the discussion to the best-constrained muon decay parameters as charged lepton flavor violating processes, such as $\mu\to e\gamma$, provide substantially more stringent constraints of the relevant operator coefficients.

Next we outlined our $\chi^2$ approach to comparing data with these theory predictions. This allowed us to provide constraints on flavor general aggregate combinations of LEFT coefficients, however due to the structure of muon decay these constraints did not allow for a straightforward power counting interpretation of the aggregate variables. For those variables which did correspond to a LEFT-like power counting, we found lower limits on the scale of new physics corresponding to a few hundred GeV to a few TeV, subject to the assumption that the UV couplings were order one.

In order to find results more consistent with the LEFT power counting we considered the case where new physics couples to all neutrinos identically, the ``flavor-democratic'' assumption. This allowed for a fit consistent with the power counting in the LEFT. We further discussed measures of how well this assumption agreed with the data, from a frequentist interpretation we found the assumption was consistent with the data. The scale of new physics inferred from this fit was consistent with those scales obtained in the flavor general fit.
 
Our next step was to discuss how new physics from above the electroweak scale results in a SMEFT interpretation which subsequently matches onto the LEFT. This resulting in a different interpretation of the power counting: dimension-six vector-like operators in the LEFT corresponded to dimension-six vector-like SMEFT operators, but the lepton number violating operators instead descend from dimension-seven operators. This immediately leads to a truncation issue, the LEFT interpretation misinterprets the importance of various contributions (subject to the SMEFT assumption). We then made the choice to calculate amplitudes to dimension-seven in the SMEFT, resulting in observables with partial results at order $1/\Lambda^6$ while neglecting dimension-eight operators contributions. We used this to motivate neglecting dimension-eight operator contributions, however stressed that this was a deliberate choice to neglect contributions of equal or lower order.

Formally this project demonstrates an excellent opportunity to demonstrate how scales are connected through EFT interpretations, from the muon mass scale a about 100 MeV to the electroweak scale $v\sim 246$ GeV, then up to some new physics scale in the potentially TeV range. However, because of the flavor structures in muon decay, the leading renormalization group dependence generally leads to only order a few percent corrections. This level of precision was negligible compared with the assumptions made elsewhere in our work (e.g. flavor assumptions), so we neglected the renormalization group dependence. This would, however, be relevant for a detailed study of a specific UV model with a specified flavor structures. It also proved necessary for discussions of lepton number violating operators which can mix into the Weinberg operator.

Finally we looked at specific UV models and how they mapped onto the SMEFT. We were then able to constrain these models directly, subject to flavor assumptions, from muon decay. Interestingly, we could formulate an effective number of neutrinos participating in the decay and constrain that quantity subject to further assumptions. In cases of a single multiplet extension of the SM we demonstrated that muon decay, occurring at around the 100 MeV scale, results in constraints on new physics typically in the TeV range, subject to an assumption of order one couplings. Unfortunately for these single multiplet extensions, any that admitted dimension-seven lepton number violation in the SMEFT also resulted in the Weinberg operator generating tree-level contributions to neutrino masses and therefore muon decay did not produce constraints yielding information beyond that obtained from neutrino mass constraints. For this reason, we went on to study two examples of two-field extensions of the SM which avoid the Weinberg operator. While muon decay measurements result in meaningful constraints on the scale of new physics for these fields, we found that RGE running resulted in the radiative generation of the Weinberg operator which were inconsistent with the upper bound on the absolute scale of neutrino masses. 

We conclude that muon decay is a useful probe of up to multi-TeV heavy physics. The different muon decay parameters allow us to look at different Lorentz and chiral structures as well as real and imaginary parts of certain combinations of parameters. When looked at from a SMEFT interpretation, the structures become much simpler, with lepton number violation occurring at a higher order. Due to the non-interference of lepton number violating amplitudes with the SM, they are suppressed as $1/\Lambda^6$ in the EFT framework. In these cases neutrino mass constraints coming from operator mixing at one-loop are generally more restrictive.

\section*{Acknowledgements}
The authors thank I. Brivio for discussions of operator counting and hermiticity. TC would like to acknowledge the Mainz Institute for Theoretical Physics (MITP) of the Cluster of Excellence PRISMA+ (Project ID 390831469) for enabling him to complete a portion of this work.

\newpage
\begin{appendix}
\numberwithin{equation}{section}

\section{Kinematics and Phase Space integrals}\label{sec:kinPS}

In this section we closely follow the textbook by Scheck \cite{Scheck:1996ur} and define the momenta as follows:
\begin{equation}
\mu_q\to \overset{\scriptscriptstyle(-)}{\nu}_{k_1}+\overset{\scriptscriptstyle(-)}{\nu}_{k_2}+e_p\, ,
\end{equation}
that is the momentum assigned to the muon is $q$, that for the electron is $p$, etc. The labels $k_1$ and $k_2$ are integrated over in deriving the Michel parameters, so the assignment to (massless) neutrinos is arbitrary. When treating tau decays, the tau momentum is also taken to be $q$.

We take $s_0=(0,P_\mu \hat n_0)$ to be the muon spin four-vector, with $P_\mu$ the muon polarization and $\vec n_0$ a unit vector in the direction of the muon spin expectation value. A similar spin vector can be defined for the electron,
\begin{equation}
s_1=\left(\frac{1}{m_e}\vec p\cdot \hat n_1, \hat n_1+\frac{\vec p\cdot \hat n_1}{m_e(E+m_e)}\vec p\right)\, ,
\end{equation}
which projects onto the electron spin state with spin along $\vec n_1$ \cite{Scheck:1996ur}.

We work in the muon rest frame, $q^\mu=(m_\mu,\vec 0)$, and express the electron four-momentum us $ p^\mu=(E_e,\vec p)$. Working out the relevant products of four-vectors we obtain:
\begin{align}
q^2&=m_\mu^2\, ,\\
p^2&=m_e^2\, ,\\
q\cdot p &= m_\mu E_e\, ,\\
q\cdot s_0 &= 0\, ,\\
q\cdot s_1 &= \frac{m_\mu}{m_e}\vec p\cdot \hat n_1\, ,\\
p\cdot s_0 &= -P_\mu\,\vec p\cdot \hat n_0\, ,\\
p\cdot s_1 &= 0\, ,\\
s_0\cdot s_1 &=-P_\mu\left(\hat n_0\cdot\hat n_1+\frac{\vec p\cdot \hat n_1\, \vec p\cdot \hat n_0}{m_e(E_e+m_e)}\right)\, .
\end{align}

The PDG defines muon decay kinematics dependent on the electron spin in terms of the product of $\hat \zeta\cdot \vec P_e$. From a theory perspective we are able to choose to align $\zeta$ with the $\hat n_1$ direction.

The product $p\cdot s_0$ is independent of the electron spin orientation:
\begin{equation}
p\cdot s_0=-P_\mu|\vec p|\cos\theta\, .
\end{equation}
Taking $\hat e_3$ to be along $\vec p$ we find $q\cdot s_1$ lies strictly along $\hat e_3=\hat p$:
\begin{eqnarray}
q\cdot s_1=\frac{m_\mu}{m_e}|\vec p|\zeta_3\, .
\end{eqnarray}
The vectors $\hat p$ and $\hat n_0$ define the decay plane. The $\hat e_1$ direction is taken to be the component of the muon spin direction orthogonal to the electron momentum,
\begin{equation}
\hat e_1=\frac{\hat n_0-(\hat n_0\cdot \hat p)\hat p}{\sqrt{1-(\hat n_0\cdot \hat p)^2}}=\frac{\hat n_0-\cos\theta\hat p}{\sin\theta}\, ,
\end{equation}
and therefore lies in the decay plane orthogonal to $\hat p$.
The product of the electron and muon spins then decomposes over the $\hat e_1$ and $\hat e_3$ directions:
\begin{equation}
s_0\cdot s_1=-P_\mu\left(\sin\theta \zeta_1+\frac{E_e}{m_e}\cos\theta\zeta_3\right)\, .
\end{equation}
Finally $\hat e_2$ is taken to be normal to the decay plane. In this case only the structure, $\hat n_0\cdot (\hat n_1\times \hat p)$ contributes, which in four vectors corresponds to:
\begin{equation}
\epsilon_{\mu\nu\rho\sigma}s_0^\mu s_1^\nu q^\rho p^\sigma=m_\mu P_\mu |\vec p|\sin\theta\zeta_2
\end{equation}

\subsection*{Phase space integrations}
In the LEFT, we need to perform phase space integrations of squared amplitudes dependent on higher orders in momentum than treated in Scheck or for example \cite{Marquez:2022bpg}. Taking $Q\equiv k_1+k_2=q-p$, and defining
\begin{equation}
\int d\Phi_2=\int \frac{d^3k_1}{2k_1^0}\frac{d^3k_2}{2k_2^0}\delta^{(4)}(Q-k_1-k_2)\, ,
\end{equation}
we obtain:
\begin{align}
\int d\Phi_2\, 1&=\frac{\pi}{2}\, ,\\
\int d\Phi_2\, k_1^\mu&=\frac{\pi}{4}Q^\mu\, ,\\
\int d\Phi_2\, k_1^\mu k_1^\nu&=\frac{\pi}{24}\left(4Q^\mu Q^\nu-Q^2\eta^{\mu\nu}\right)\, ,\\
\int d\Phi_2\, k_1^\mu k_1^\nu k_1^\rho&=\frac{\pi}{48}\left[6Q^\mu Q^\nu Q^\rho-Q^2\left(Q^\mu \eta^{\nu\rho}+Q^\nu \eta^{\mu\rho}+Q^\rho\eta^{\mu\nu}\right)\right]\, .
\end{align}
It is important to note that all momenta in these basis integrands are the same, i.e. they are all $k_1$. These results also implicitly assume massless neutrinos. When applying these identities to our results we simply substitute $k_2=Q-k_1$ so that our integrals depend only on $k_1$. 

The decay width for the muon is then given as:
\begin{equation}
\frac{d^2\Gamma}{dE_e d\cos\theta}=\frac{1}{4m_\mu}\frac{\sqrt{E_e^2-m_e^2}}{(2\pi)^5}\int d\phi\, d\Phi_2|\, \mathcal M|^2\, ,
\end{equation}
with $\mathcal M$ the amplitude for the process. As our definition of the two body phase space excludes factors of $2\pi$ some additional factors appear explicitly in our expression for the differential decay width. For identical neutrinos in the final state an additional factor of one half is necessary.

Defining the maximum electron energy to be,
\begin{equation}
W = \frac{1}{2m_\mu}(m_\mu^2+m_e^2)\, ,
\end{equation}
we make a change of variables to $x\equiv E_e/W$ giving:
\begin{equation}
\frac{d^2\Gamma}{dx d\cos\theta}=\frac{W}{4m_\mu}\frac{\sqrt{W^2x^2-m_e^2}}{(2\pi)^5}\int d\phi\, d\Phi_2|\, \mathcal M|^2\, .
\end{equation}

\section{Michel ``basis functions''}\label{sec:michelbasis}

We define the following set of basis functions to aid in extracting the Michel parameters. Starting with electron polarization independent Michel parameters we define:
\begin{eqnarray}
\phi_1(x,x_0,\cos\theta)&=&\sqrt{x^2-x_0^2}x(1-x)\, ,\\
\phi_2(x,x_0,\cos\theta)&=&\sqrt{x^2-x_0^2}\frac{2}{9}(4x^2-3x-x_0^2)\, ,\\
\phi_3(x,x_0,\cos\theta)&=&\sqrt{x^2-x_0^2}x_0(1-x)\, ,\\
\phi_4(x,x_0,\cos\theta)&=&\cos\theta(x^2-x_0^2)\frac{1}{3}(1-x)\, ,\\
\phi_5(x,x_0,\cos\theta)&=&\cos\theta(x^2-x_0^2)\frac{2}{9}\left[(4x-3)+\sqrt{1-x_0^2}-1\right]\, .
\end{eqnarray}

These correspond to the kinematic structures multiplying the Michel parameters in the differential decay rate. For example, considering only $\rho$ dependence from the decay rate we have:
\begin{equation}
\frac{d^2\Gamma}{dxd\cos\theta}\sim\rho \frac{m_\mu}{4\pi^3}W^4G_F^2\sqrt{x^2-x_0^2}\frac{2}{9}(4x^2-3x-x_0^2)\, .
\end{equation}
Defining the Gram matrix,
\begin{equation}
G_{ij}=\int_{x_0}^1dx\, \int_{-1}^1d\cos\theta\phi_i\phi_j\, ,
\end{equation}
we can solve for the Michel parameters by taking:
\begin{equation}
\begin{pmatrix}
A\equiv \frac{G_F^2 m_\mu W^4}{4\pi^3}\\
A\rho\\
A\eta\\
\pm P_\mu A \xi\\
\pm P_\mu A \xi\delta
\end{pmatrix}=G_{ij}^{-1}b_j\, ,\label{eq:solveforMichel}
\end{equation}
where we have defined:
\begin{equation}
b_j=\left(\int_{x_0}^1dx\, \int_{-1}^{1}d\cos\theta \frac{d^2\Gamma}{dxd\cos\theta}\phi_j\right)\, .
\end{equation}
For muon decay $\pm1$ should be understood to be $+1$ for positive muon decay, and $-1$ for negative muon decay.

The same approach is taken for the other Michel parameters. We split these cases by their projection onto our choice of $\zeta_i$ as mentioned in \Cref{sec:kinPS}.

\subsection*{$\zeta_1$}
The $\zeta_1$ projection corresponds to $F_{T_1}$ in the PDG notation \cite{ParticleDataGroup:2024cfk}. The basis functions we choose are:
\begin{eqnarray}
\phi_{\zeta_1,1}(x,x_0,\cos\theta)&=&\sqrt{x^2-x_0^2}\sqrt{1-\cos^2\theta}(1-x)x_0\, ,\\
\phi_{\zeta_1,2}(x,x_0,\cos\theta)&=&\sqrt{x^2-x_0^2}\sqrt{1-\cos^2\theta}(x^2-x_0^2)\, ,\\
\phi_{\zeta_1,3}(x,x_0,\cos\theta)&=&\sqrt{x^2-x_0^2}\sqrt{1-\cos^2\theta}(4x-3x^2-x_0^2)\, .
\end{eqnarray}
These basis functions allow for the extraction of $\xi''$, $\eta$, and $\eta''$ respectively. The simultaneous extraction of $\eta$ from $\phi_{\zeta_1,2}$ and $\phi_3$ provides a useful consistency check of our results.

\subsection*{$\zeta_2$}
The $\zeta_2$ projection corresponds to $F_{T_2}$ in the PDG notation \cite{ParticleDataGroup:2024cfk}. The basis functions we choose are:
\begin{eqnarray}
\phi_{\zeta_2,1}(x,x_0,\cos\theta)&=&(x^2-x_0^2)\sqrt{1-\cos^2\theta}(1-x)\, ,\\
\phi_{\zeta_2,2}(x,x_0,\cos\theta)&=&(x^2-x_0^2)\sqrt{1-\cos^2\theta}\sqrt{1-x_0^2}\, .
\end{eqnarray}
These basis functions allow for the extraction of $\alpha'/A$ and $\beta'/A$ respectively.

\subsection*{$\zeta_3$}
The $\zeta_3$ projection corresponds to $F_{IP}$ and $F_{AP}$ in the PDG notation \cite{ParticleDataGroup:2024cfk}. The basis functions we choose are:
\begin{eqnarray}
\phi_{\zeta_3,1}(x,x_0,\cos\theta)&=&(x^2-x_0^2)(2-2x+\sqrt{1-x_0^2})\, ,\\
\phi_{\zeta_3,2}(x,x_0,\cos\theta)&=&(x^2-x_0^2)(4x-4+\sqrt{1-x_0^2})\,,\\
\phi_{\zeta_3,3}(x,x_0,\cos\theta)&=&\cos\theta\sqrt{x^2-x_0^2}(2x^2-x-x_0^2)\, ,\\
\phi_{\zeta_3,4}(x,x_0,\cos\theta)&=&\cos\theta\sqrt{x^2-x_0^2}(4x^2-3x-x_0^2)\, ,\\
\phi_{\zeta_3,5}(x,x_0,\cos\theta)&=&\cos\theta\sqrt{x^2-x_0^2}(1-x)x_0\, .
\end{eqnarray}
These basis functions allow for the extraction of $\xi'$, $\delta$, $\xi''$, $\rho$, and $\eta''$. Again, redundancy with previously extracted Michel parameters provides a useful cross check of our results.

\section{Hermiticity of vector-like LEFT operators}\label{app:hermiticity}
Operators of the form,
\begin{equation}
\mathcal O^{prst} = (\bar\psi^p\gamma^\mu P_\pm\psi^r)(\bar\chi^s\gamma_\mu P'_\pm\chi^t)\, ,\label{eq:OprstApp}
\end{equation}
are often referred to as hermitian in theories like the LEFT and SMEFT. Here $P'_\pm$ is written with a prime just to indicate its sign need not be the same as that in the preceding bilinear. Taking the hermitian conjugate of the operator, however, gives:
\begin{equation}
\left(\mathcal O^{prst}\right)^\dagger=(\bar\psi^r\gamma^\mu P_\pm\psi^p)(\bar\chi^t\gamma_\mu P'_\pm\chi^s)\, ,
\end{equation}
which is not equal to the expression of \Cref{eq:OprstApp}. When we include the Wilson coefficients and an implicit sum over the flavor indices we have,
\begin{equation}
\left[c^{prst}\mathcal O^{prst}\right]^\dagger=\left[c^{prst}(\bar\psi^p\gamma^\mu P_\pm\psi^r)(\bar\chi^s\gamma_\mu P'_\pm\chi^t)\right]^\dagger=(c^*)^{prst}(\bar\psi^r\gamma^\mu P_\pm\psi^p)(\bar\chi^t\gamma_\mu P'_\pm\chi^s)\, .
\end{equation}
At this stage we impose hermiticity of the effective Lagrangian. In the conventional operator bases used in the LEFT and SMEFT, hermitian conjugates which belong to the same flavor-indexed operator class are not treated as independent operators \cite{Henning:2015alf,Lehman:2015via}. Relabelling the indices in the last expression such that the operator again has the ordering $prst$, hermiticity then implies:
\begin{equation}
(c^{prst})^*= c^{rpts}\, .\label{eq:WChermitianassocApp}
\end{equation}
This allows us to re-express this relation as,
\begin{equation}
\left[c^{prst}\mathcal O^{prst}\right]^\dagger=c^{prst}\mathcal O^{prst}\, ,
\end{equation}
and therefore the sum over flavors of the operator multiplied by its flavor-dependent Wilson coefficients is hermitian. The final equality should therefore always be understood together with the implicit Wilson coefficient constraint in \Cref{eq:WChermitianassocApp}. As such, referring to an \textit{operator} of this form as hermitian, is shorthand for the discussion in this appendix. This is implicitly used across the LEFT and SMEFT literature when referring to (flavor dependent) operators as hermitian.

\end{appendix}

\newpage
\bibliography{bibtex.bib}


\end{fmffile}
\end{document}